\documentclass[seceq]{ptptex}
\usepackage{graphicx}
\usepackage{wrapft}
\usepackage{bm}
\def\mbf#1{\hbox{\boldmath $#1$}}
\def\eq#1{Eq.\ (\ref{#1})} 
\def\bN{{\mbf N}}

\def\bk{{\mbf k}}
\def\bp{{\mbf p}}
\def\bq{{\mbf q}}
\def\bz{{\mbf z}}
\def\CK{{\cal K}}
\def\CP{{\cal P}}
\def\H{{\scriptstyle \frac{1}{2}}}
\def\3H{{\scriptstyle \frac{3}{2}}}
\notypesetlogo 
\title{Quark-Model Baryon-Baryon Interaction Applied to Neutron-Deuteron Scattering. I}
\subtitle{Noyes-Kowalski Approach to the AGS Equations}
\author{Yoshikazu \textsc{Fujiwara} and Kenji \textsc{Fukukawa}}
\inst{Department of Physics, Kyoto University, Kyoto 606-8502, Japan}
\recdate{July 13, 2010; Revised August 11, 2010}

\abst{We solve the neutron-deuteron ($nd$) scattering in the Faddeev formalism,
employing the $NN$ sector of the quark-model baryon-baryon interaction fss2.
The energy dependence of the $NN$ interaction, inherent to the resonating-group formulation of the quark-model
baryon-baryon interaction, is eliminated by the standard off-shell transformation utilizing the normalization
kernel of two-cluster systems. An extra nonlocality originating from this procedure
is very important to reproduce all the $nd$ scattering observables below $E_{n} \leq 65$ MeV. 
A new algorithm to solve the Alt-Grassberger-Sandhas (AGS) equations is presented in the Noyes-Kowalski method.
The treatment of the moving singularity from the three-body Green function is shown in detail,
which is based on the spline function method originally developed by the Bochum-Krakow group.
The predicted elastic differential cross sections reproduce the observed deep cross section minima at
$E_{n}=35\hbox{--}65$ MeV and $ \theta _{\rm cm}=130^{\circ}\hbox{--}135^{\circ}$,
which is consistent with the nearly correct triton binding energy predicted by fss2 without the three-body force.}

\PTPindex{205}  

\begin{document}
\maketitle
\section{Introduction}
The QCD-inspired spin-flavor $SU_{6}$ quark model (QM) for the baryon-baryon interaction,
developed by the Kyoto-Niigata group, is a unified model describing interactions between full octet-baryons.\cite{PPNP}
It is given by the Born kernel formulated in the resonating-group method (RGM) for interacting three-quark clusters.
The short-range part is described by an effective one-gluon exchange,
while the medium- and long-range parts are dominated by meson-exchange potentials between quarks.
The model parameters are constrained to reproduce all the two-nucleon data
and available low-energy hyperon-nucleon scattering data.
These QM baryon-baryon interactions are characterized by the nonlocality and energy dependence
inherent to the RGM framework. For example, the Pauli-forbidden state appears on the quark level
in certain channels of the strangeness sector, as a result of the exact antisymmetrization of six quarks.
The short-range repulsion described by the nonlocality of the quark-exchange kernel
gives quite different off-shell properties from the standard meson-exchange potentials.
The energy dependence of the interaction is eliminated by the standard off-shell transformation,
utilizing a square root of the normalization kernel $N$.\cite{Sa73,Sa77,FW82,renRGM}
This procedure yields an extra nonlocality,
whose effect was examined in detail for the three-nucleon ($3N$) bound state and for the hypertriton.\cite{ren}
The advantage of the larger triton binding energy caused by the QM nucleon-nucleon ($NN$) interaction,
namely, the deficiency of 350 keV, predicted using the most recent model fss2,
is still much smaller than the standard values of 0.5 -- 1 MeV,\cite{No00}
given by the modern meson-exchange $NN$ potentials.
It is therefore interesting to examine predictions based on the QM $NN$ interaction for the $3N$ scattering,
especially in this renormalized framework without the explicit energy dependence of the RGM kernel.

An attempt to investigate the effect of nonlocality of the underlying two-body interaction on three-cluster systems
composed of composite particles is actually not new. Even if we restrict our interest to the $3N$ system,
there are many investigations to improve the triton binding energy and the neutron-deuteron ($nd$) scattering
observables, from various types of nonlocal $NN$ interaction.\cite{Do03,Vi06,Ta92,Do08}.
There is, however, no extensive investigation of the $3N$ system,
employing a consistent $NN$ nonlocal interaction that reproduces all the two-nucleon data with satisfactory accuracy.
 
In this series of papers, we apply our QM $NN$ interaction fss2 to the $nd$ scattering in the Faddeev formalism
for systems of composite particles \cite{TRGM,RED,OCMFAD}.
The Alt-Grassberger-Sandhas (AGS) equations \cite{AGS} are solved in the momentum representation,
using the off-shell RGM $t$-matrix obtained from the energy-independent renormalized RGM kernel.
The Gaussian nonlocal potential constructed from the fss2 is used in the isospin basis.\cite{apfb08}
The singularity of the $NN$ $t$-matrix from the deuteron pole is handled by the Noyes-Kowalski method.\cite{NO65,KO65}
To the best of our knowledge, this simple method to solve a singular Lippmann-Schwinger equation
has never been seriously applied to the AGS equation, despite the fact that this method
was originally developed for application to three-body problems.
We will show a new practical algorithm to solve the AGS equations in this Noyes-Kowalski method.
Another notorious moving singularity of the three-body Green function for the free motion is treated by
the standard spline interpolation technique developed by the Bochum-Krakow group.\cite{spline82,Wi03,PREP,Liu05}
Here again, we will give a detailed procedure and formulations since they do not seem to be available in the literature.
We mainly use the channel-spin formalism, which is convenient to discuss the $nd$ scattering.
The accuracy of the numerical calculations can be checked by examining the optical theorem for the $nd$ scattering.
The $NN$ interaction up to $I_{\rm max}=4$ with a sufficient number of partial waves of the three-body system is included,
resulting in the maximum neutron incident energy of about $E_{n}=65$ MeV in the laboratory system.
Here, $I_{\rm max}$ is the maximum value of the two-nucleon angular momentum included in the calculation.
In this paper, we only show the results of total and differential cross sections of the $nd$ elastic scattering.
Polarizations and deuteron breakup cross sections are discussed in subsequent papers.\footnote{
A preliminary report on the present subject is found in Refs.\,\citen{fb19,KF10,FK10}.}
The details of the Gaussian nonlocal potentials and an application to the $S$-wave $nd$ scattering length
will be given in separate papers.\cite{SCL10}

The organization of this paper is as follows. In \S 2.1, we first recapitulate our QM baryon-baryon interaction
and then outline the whole procedure to solve AGS equations for the $nd$ scattering.
An emphasis is put on how to calculate an extra nonlocal kernel
originating from the elimination of the energy dependence in the RGM kernel.
The essential part of this analytic formulation is given in Appendix A.
In \S 2.2, we exhibit a new formulation of AGS equations,
incorporating the deuteron singularity of the $NN$ $t$-matrix in the Noyes-Kowalski method.
The basic equation is convenient to derive the optical theorem by a simple manipulation in the operator formalism.
The contribution of the breakup cross sections to the optical theorem is thoroughly discussed in \S 2.3.
Practical calculations are made by the procedure given in \S 2.4,
where the moving singularity of the three-body Green function for the free motion is processed by
the subtraction method and the spline interpolation technique.
The detailed procedure to calculate the basic integral $Q_{k \mu \nu}$ is given in Appendix B.
The total cross sections derived from the optical theorem and $nd$ elastic differential cross sections
are shown in \S 3.1 and \S 3.2, respectively. It is found that the predicted elastic differential cross sections
reproduce the observed deep cross section minima on the high-energy side,
which is consistent with the nearly correct triton binding energy predicted using fss2 without the three-body force.
The last section is devoted to a summary.

\section{Formulation}
\subsection{Quark-model baryon-baryon interaction}
The QM baryon-baryon interaction is a low-energy effective model
that introduces some of the essential features of QCD (quantum chromodynamics) characteristics.
The color degree of freedom of quarks is explicitly incorporated
into the nonrelativistic spin-flavor $SU_{6}$ quark model,
and the full antisymmetrization of quarks is carried out in the RGM formalism.
The gluon exchange effect is represented in the form of the quark-quark interaction.
The confinement potential is a phenomenological $r^2$-type potential,
which has a favorable feature in that it does not contribute to the baryon-baryon interactions.
We use a color analogue of the Fermi-Breit (FB) interaction,
motivated from the dominant one-gluon exchange process in the high-momentum region.
We postulate that the short-range part of the baryon-baryon interaction is well described by the quark degree of freedom.
This includes the short-range repulsion and spin-orbit force,
both of which are successfully described by the FB interaction.
On the other hand, the medium-range attraction and long-range tensor force, especially afforded by the pions,
are extremely nonperturbative. These are therefore most relevantly described by
the effective meson exchange potentials (EMEPs). The most recent model fss2 includes the vector-meson exchange EMEP, 
in addition to the scalar- and pseudoscalar-meson exchange potentials.\cite{FU02a,FU01b}

The quark-model Hamiltonian $h$ consists of the phenomenological confinement potential $U^{\rm Cf}_{ij}$,
the colored version of the full FB interaction $U^{\rm FB}_{ij}$ with explicit quark-mass dependence,
and the EMEP $U^{\Omega}_{ij}$ generated from the scalar ($ \Omega $=S), pseudoscalar (PS), and vector (V)
meson exchange potentials acting between quarks:
\begin{eqnarray}
h=\sum^{6}_{i=1}\left(m_{i}c^{2}+\frac{\bp^{2}_{i}}{2m_{i}}-T_{G}\right)
+\sum^{6}_{i<j} \left(U^{\rm Cf}_{ij}+U^{\rm FB}_{ij}
+U^{\rm S}_{ij}+U^{\rm PS}_{ij}+U^{\rm V}_{ij} \right)\ . 
\label{qm1}
\end{eqnarray}
The RGM equation for the relative-motion wave function $ \chi({\bm r})$ reads
\begin{eqnarray}
\langle \phi(3q)\phi(3q)\vert E-h \vert{\cal A}\left \{\phi(3q)\phi(3q)\chi({\bm r})\right \}\rangle=0\ ,
\label{qm2}
\end{eqnarray}
where $ \phi(3q)$ is a simple harmonic-oscillator (h.o.)~shell-model wave function for the three-quark clusters. 
We solve this RGM equation in the momentum representation.\cite{LSRGM}
If we rewrite the RGM equation in the form of the Schr{\"o}dinger-type equation as
$[\varepsilon-h_{0}-V_{\rm RGM}(\varepsilon)]\chi({\bm r})=0$,
the potential term, $V_{\rm RGM}(\varepsilon)=V_{\rm D}+G+\varepsilon K$, becomes nonlocal and energy-dependent.
Here, $V_{\rm D}$ represents the direct potential of EMEPs,
$G$ includes all the exchange kernels for the interaction and kinetic-energy terms,
$K=1-N$ is the exchange normalization kernel, and $ \varepsilon $ is the total energy in the center-of-mass (cm)~system,
measured from the two-cluster threshold. We calculate the plane-wave matrix elements of $V_{\rm RGM}(\varepsilon)$,
and set up with the Lippmann-Schwinger equation for the RGM $t$-matrix.
This approach is convenient to proceed to the $G$-matrix calculations \cite{KO00,GRGM}
and to the Faddeev calculations with some special considerations of the Pauli-forbidden states.\cite{TRGM,RED,OCMFAD}

The energy-independent renormalized RGM kernel $V^{\rm RGM}$ for a two-cluster system is given by~\cite{renRGM}
\begin{eqnarray}
V^{\rm RGM}=V_{\rm D}+G+W\ .  \label{qm3}
\end{eqnarray}
The nonlocal kernel $W$ appears through the elimination of the energy dependence, and is given by
\begin{eqnarray}
W=\Lambda \frac{1}{\sqrt{N}}h \frac{1}{\sqrt{N}}\Lambda-h\ .
\label{qm4}
\end{eqnarray}
Here, $N=1-K$ is the normalization kernel, $h=h_{0}+V_{\rm D}+G$ with $h_{0}$ being the kinetic energy
for the two-cluster relative motion, and $ \Lambda=1-|u \rangle \langle u|$ is the two-cluster Pauli projection operator,
where $|u \rangle $ is a Pauli-forbidden state satisfying $K|u \rangle=|u \rangle $.
The Born kernel $W(\bp, \bp')$ of \eq{qm4}, or its partial wave component, is calculated accurately
using an analytical procedure to obtain $\CK=\Lambda (1/\sqrt{N}-1)\Lambda $ in the momentum representation.
This is discussed in Appendix A. In the $NN$ sector, no Pauli-forbidden state appears on the quark level,
so that we can simply set $ \Lambda=1$ in the following formulations.
An advantage of using $V^{\rm RGM}$ is that the two-cluster RGM equation takes the form
of the usual Schr{\"o}dinger equation in the Pauli-allowed model space,
and the relative wave function is properly normalized.
This Schr{\"o}dinger-type equation for the relative wave function 
gives the same asymptotic behavior as the original RGM equation,
thus preserving the phase shifts and physical observables for the two-cluster systems.
The difference between the previous energy-dependent RGM kernel,
$V_{\rm RGM}(\varepsilon)=V_{\rm D}+G+\varepsilon K$,
and $V^{\rm RGM}$ in \eq{qm3} is essentially a replacement of $ \Lambda(\varepsilon K)\Lambda $ with $W$.
The value of $ \varepsilon $ is, however, not properly defined in the three-cluster system, in particular,
for the scattering systems. In the following, we will consistently use the energy-independent renormalized RGM kernel
$V^{\rm RGM}$ in \eq{qm3}, both for the bound-state solution and the scattering problems.

The three-cluster equation for the $3N$ bound state is written as
\begin{eqnarray}
\left[\,E-H_{0}-V^{\rm RGM}_{\alpha}-V^{\rm RGM}_{\beta}-V^{\rm RGM}_{\gamma}\,\right] \Psi=0\ ,
\label{qm5}
\end{eqnarray}
where $ \alpha $, $ \beta $, and $ \gamma $ denote three independent pairs of two-cluster subsystems,
$H_{0}$ is the free three-body kinetic-energy operator, and $V^{\rm RGM}_{\alpha}$ stands for the RGM kernel
in \eq{qm3} for the $ \alpha $-pair, etc.
In Ref.~\citen{ren}, we have solved \eq{qm5} in the Faddeev formalism
\begin{eqnarray}
\psi=G_{0}tP \psi \ ,
\label{qm6}
\end{eqnarray}
where $G_{0}=G_{0}(z)=1/(z-H_{0})$ is the three-body Green function for the free motion,
$P=P_{(12)}P_{(13)}+P_{(13)}P_{(12)}$ is a permutation operator for the rearrangement,
and $t=t(z-\bar{h}_{0})$ is the $NN$ $t$-matrix derived by solving the Lippmann-Schwinger equation
$t=v+vG_{0}t$ with $v=V^{\rm RGM}$ in \eq{qm3}. Here, $z=E$ is the total energy,
which is below the deuteron energy $ \varepsilon _{d}~(< 0)$ for the $3N$ bound state.
The energy argument of $t$ is $z-\bar{h}_{0}$, where $ \bar{h}_{0}$ is the kinetic-energy operator
for the relative momentum $\bq=(2\bk_{3}-\bk_{1}-\bk_{2})/3$.
Another momentum for the two-nucleon relative motion is denoted by $\bp=(\bk_{1}-\bk_{2})/2$
with $h_{0}$ being the corresponding kinetic-energy operator.
The three-body kinetic-energy operator is therefore $H_{0}=h_{0}+\bar{h}_{0}$.
In \eq{qm6}, $ \psi $ is the Faddeev component that yields the total wave function $ \Psi $ in \eq{qm5}
through $ \Psi=(1+P)\psi $. The basic equation for the $nd$ scattering is the AGS equation,
which can be expressed as \cite{PREP}
\begin{eqnarray}
U|\phi \rangle={G_{0}}^{-1} P|\phi \rangle+P\,t\,G_{0}\,U|\phi \rangle \ ,
\label{qm7}
\end{eqnarray}
where $|\phi \rangle=|\bq_{0},\psi _{d}\rangle $ is the plane-wave channel wave function
with $|\psi _{d}\rangle $ being the deuteron wave function. 
In this case, $z=E+i0$ is the total energy approached from the upper side 
of the real axis in the complex energy plane, 
and $E=E_{\rm cm}+\varepsilon _{d}$ with $E_{\rm cm}=(3 \hbar^{2}/4M){q_{0}}^{2}$
being the neutron incident energy in the cm system.
We use an average nucleon mass $M=(M_{p}+M_{n})/2$ in the isospin formalism.
The scattering amplitude for the elastic scattering is obtained from $ \langle \phi|U|\phi \rangle $.
It is important that \eq{qm7} also provides information on the full breakup process of the deuteron.
The transition amplitude for the breakup process is given by
\begin{eqnarray}
U_{0}|\phi \rangle=(1+P)tG_{0}U|\phi \rangle=(1+P)T|\phi \rangle \ ,
\label{qm8}
\end{eqnarray}
where $T=tG_{0}U$ corresponds to the three-body $t$-matrix. The breakup cross sections are obtained from the amplitude
$ \langle \bp \bq |U_{0}|\phi \rangle $ with the corresponding energy $E=(\hbar^{2}/M)(\bp^{2}+(3/4)\bq^{2})$. 

\subsection{Noyes-Kowalski method for the singularity of the $NN$ $t$-matrix}
For the description of the $nd$ elastic scattering, it is convenient to use the channel-spin representation,
in which the angular-spin functions are defined through
\begin{eqnarray}
& & |\bp, \bq;123 \rangle=\sum _{\gamma}|p,q,\gamma \rangle~\langle \gamma|\widehat{\bp},\widehat{\bq};123 \rangle \,\nonumber \\
& & \langle \widehat{\bp},\widehat{\bq};123|\gamma \rangle=\left[Y_{\ell}(\widehat{\bq})~\bigl[
\bigl[Y_{\lambda}(\widehat{\bp})\chi _{st}(1,2)\bigr]_{I}~\chi_{\H \H}(3)\bigr]_{S_{c}}\right]_{JJ_{z};\H T_{z}}\ ,
\label{fm7}
\end{eqnarray}
with $ \gamma=\ell[(\lambda s)I \H]S_{c};JJ_{z};(t \H)\H T_{z}$.
Here, $ \chi _{st}(1,2)$ etc. are the spin-isospin wave functions.
In particular, the deuteron channels are specified by
\begin{eqnarray}
\gamma _{d}=\ell[(\lambda 1)1 \H]S_{c};JJ_{z};(0 \H)\H T_{z}\ ,
\label{fm8}
\end{eqnarray}
with $ \lambda=0$ and 2, corresponding to the $S$-wave and $D$-wave components, respectively.
We use $t=\sum _{\gamma}|\gamma \rangle t_{\gamma}\langle \gamma|$, and separate the $ \gamma $-sum into
$ \gamma \notin (\gamma _{d})$ and $ \gamma \in (\gamma _{d})$.
The $t$-matrix of the deuteron channel has the deuteron pole, which we explicitly separate as
\begin{eqnarray}
t_{\gamma _{d}}=\frac{\widehat{t}_{\gamma _{d}}}{z-\bar{h}_{0 \gamma _{d}}-\varepsilon _{d}}\ .
\label{fm9}
\end{eqnarray}
If we use the completeness relationship, $ \int^{\infty}_{0}q^{2}\,d\,q~|q \rangle \langle q|=1$,
and the spectral decomposition of the two-nucleon Green function, we can easily show
\begin{eqnarray}
\ \hspace{-10mm} 
\frac{\widehat{t}_{\gamma _{d}}}{z-\bar{h}_{0 \gamma _{d}}-\varepsilon _{d}}
=\frac{4M}{3 \hbar^{2}}\CP \int^{\infty}_{0}q^{2}\,d\,q~|q \rangle
\frac{\widehat{t}_{\gamma _{d}}[q]}{{q_{0}}^{2}-q^{2}}\langle q|-ic \left(|q_{0}\rangle
\widehat{t}_{\gamma _{d}}[q_{0}]\langle q_{0}|\right)\ ,
\label{fm10}
\end{eqnarray}
where we have used a notation
\begin{eqnarray}
& & c=2 \pi \frac{q_{0}M}{3 \hbar^{2}}\ ,\nonumber \\
& & \widehat{t}_{\gamma _{d}}[q]=\widehat{t}_{\gamma _{d}}\left(\frac{3 \hbar^{2}}{4M}({q_{0}}^{2}-q^{2})
+\varepsilon _{d}\right) \ ,\nonumber \\
& & \widehat{t}_{\gamma _{d}}[q_{0}]=\widehat{t}_{\gamma _{d}}(\varepsilon _{d})
={g_{0}(\varepsilon _{d})}^{-1}|\psi _{d}\rangle \langle \psi _{d}|{g_{0}(\varepsilon _{d})}^{-1} \ ,
\label{fm11}
\end{eqnarray}
with ${g_{0}(\varepsilon _{d})}^{-1}|\psi _{d}\rangle=(\varepsilon _{d}-h_{0})|\psi _{d}\rangle=v|\psi _{d}\rangle $
being the deuteron solution. For the separable deuteron residue, we in fact need to make a distinction
between the $S$-wave and $D$-wave channels by writing $ \widehat{t}_{\gamma _{d}}(\varepsilon _{d})$ as
\begin{eqnarray}
& & \widehat{t}_{\gamma _{d},\widetilde{\gamma}_{d}}(\varepsilon _{d})={g_{0}(\varepsilon _{d})}^{-1}
|\psi _{\gamma _{d}}\rangle \langle \psi _{\widetilde{\gamma}_{d}}|{g_{0}(\varepsilon _{d})}^{-1}\ , \nonumber \\ [2mm]
& & \langle \psi _{d}|\psi _{d}\rangle=\sum _{\gamma _{d}} \langle \psi _{\gamma _{d}}|\psi _{\gamma _{d}}\rangle=1\ ,
\qquad \sum _{\gamma _{d}}|\gamma _{d}\rangle|\psi _{\gamma _{d}}\rangle=|\psi _{d}\rangle \ .
\label{fm12}
\end{eqnarray}
In \eq{fm10}, we should note
\begin{eqnarray}
|q_{0}\rangle \widehat{t}_{\gamma _{d}}(\varepsilon _{d})\langle q_{0}|&=&{g_{0}(\varepsilon _{d})}^{-1}
|q_0,\psi _{d}\rangle \langle q_{0},\psi _{d}|{g_{0}(\varepsilon _{d})}^{-1}
={G_{0}}^{-1}|\phi \rangle \langle \phi|{G_{0}}^{-1}\ ,
\label{fm13}
\end{eqnarray}
since
\begin{eqnarray}
{g_{0}(\varepsilon _{d})}^{-1}|q_{0},\psi_{\gamma _{d}} \rangle
&=& (\varepsilon _{d}-h_{0})|q_{0},\psi _{d}\rangle
=\left(\frac{3 \hbar^{2}}{4M}{q_{0}}^{2}+\varepsilon _{d}-h_{0}-\bar{h}_{0}\right)|q_{0},\psi _{d}\rangle \nonumber \\
&=& {G_{0}}^{-1}|q_{0},\psi _{d}\rangle={G_{0}}^{-1}|\phi \rangle \ ,
\label{fm14}
\end{eqnarray}
with $|\phi \rangle=|q_{0},\psi _{d}\rangle $. In fact, a more rigorous expression is
\begin{eqnarray}
& & \sum _{\gamma _{d},\widetilde{\gamma}_{d}}|\gamma _{d}\rangle|q_{0}\rangle 
\widehat{t}_{\gamma _{d},\widetilde{\gamma}_{d}}(\varepsilon _{d})\langle q_{0}|\langle \widetilde{\gamma}_{d}|
=\sum _{\gamma _{d},\widetilde{\gamma}_{d}}|\gamma _{d}\rangle {G_{0}}^{-1}|q_{0},\psi _{\gamma _{d}}\rangle
\langle q_{0},\psi _{\widetilde{\gamma}_{d}}|{G_{0}}^{-1}\langle \widetilde{\gamma}_{d}| \nonumber \\
& & ={G_{0}}^{-1}\sum _{\gamma _{d}}|q_{0},\gamma _{d}\rangle|\psi _{\gamma _{d}}\rangle
\sum _{\widetilde{\gamma}_{d}}\langle \psi _{\widetilde{\gamma}_{d}}|\langle q_{0},\widetilde{\gamma}_{d}|{G_{0}}^{-1}
\nonumber \\
& & ={G_{0}}^{-1}|q_{0},\psi _{d}\rangle \langle q_{0},\psi _{d}|{G_{0}}^{-1}
={G_{0}}^{-1}|\phi \rangle \langle \phi|{G_{0}}^{-1}\ .
\label{fm15}
\end{eqnarray}
Here, $ \widetilde{\gamma}_{d}$ is defined from $ \gamma _{d}$ in \eq{fm8}
by just replacing $ \lambda $ with $ \widetilde{\lambda}$.
Since this tensor coupling is cumbersome in the complicated expressions, we shall use in the following
a simple notation to represent $ \widehat{t}_{\gamma _{d},\widetilde{\gamma}_{d}}$ by $ \widehat{t}_{\gamma _{d}}$.
After all, we have obtained
\begin{eqnarray}
\sum _{\gamma _{d}}|\gamma _{d}\rangle \frac{\widehat{t}_{\gamma _{d}}}{z-\bar{h}_{0 \gamma _{d}}-\varepsilon _{d}}
\langle \gamma _{d}|
&=& \frac{4M}{3 \hbar^{2}}\sum _{\gamma _{d}}~\CP \int^{\infty}_{0}q^{2}\,d\,q~|q,\gamma _{d}\rangle
\frac{\widehat{t}_{\gamma _{d}}[q]}{{q_{0}}^{2}-q^{2}}\langle q,\gamma _{d}| \nonumber \\
& & -ic{G_{0}}^{-1}|\phi \rangle \langle \phi|{G_{0}}^{-1}\ .
\label{fm16}
\end{eqnarray}
We write the first term of the right-hand side (r.h.s.) as $ \widetilde{t}_{\gamma _{d}}$
and generalize the expression to all the channels:
\begin{eqnarray}
t_{\gamma}=\left \{\begin{array}{c}
\tilde{t}_\gamma \\ [2mm]
\widetilde{t}_{\gamma_d}-ic{G_{0}}^{-1}|\phi \rangle \langle \phi|{G_{0}}^{-1} \\
\end{array} \right.
\qquad \hbox{for} \quad
\begin{array}{c}
\gamma \notin (\gamma _{d}) \\ [2mm]
\gamma \in (\gamma _{d}) \\
\end{array} \ .
\label{fm17}
\end{eqnarray}
This yields a separation of the two-nucleon singularity from the full $t$-matrix:
\begin{eqnarray}
t=\widetilde{t}-ic{G_{0}}^{-1}|\phi \rangle \langle \phi|{G_{0}}^{-1}\ ,
\label{fm18}
\end{eqnarray}
where $ \widetilde{t}=\sum _{\gamma}|\gamma \rangle \widetilde{t}_{\gamma}\langle \gamma|$
($=\sum _{\gamma \notin(\gamma _{d})}|\gamma \rangle t_{\gamma}\langle \gamma|
+\sum _{\gamma _{d}}|\gamma _{d}\rangle \widetilde{t}_{\gamma _{d}}\langle \gamma _{d}|$) with
\begin{eqnarray}
\widetilde{t}_{\gamma _{d}}=\frac{4M}{3 \hbar^{2}}~\CP \int^{\infty}_{0}q^{2}\,d\,q~|q,\gamma _{d}\rangle
\frac{\widehat{t}_{\gamma _{d}}[q]}{{q_{0}}^{2}-q^{2}}\langle q,\gamma _{d}|\ .
\label{fm19}
\end{eqnarray}
The difference between $t$ and $ \widetilde{t}$ appears only when $q=q_{0}$ in the deuteron channel,
and $ \widetilde{t}$ also satisfies the same basic $t$-matrix equation $ \widetilde{t}=v+vG_{0}\widetilde{t}$.
If we use \eq{fm18} in \eq{qm7}, it becomes
\begin{eqnarray}
U|\phi \rangle &=& {G_{0}}^{-1}P|\phi \rangle \left[1- ic \langle \phi|U|\phi \rangle \right]
+P \widetilde{t}G_{0}U|\phi \rangle \ ,
\label{fm20}
\end{eqnarray}
since $P{G_{0}}^{-1}={G_{0}}^{-1}P$. If we define $U'|\phi \rangle $ by
\begin{eqnarray}
U|\phi \rangle = U'|\phi \rangle \left[1-ic \langle \phi|U|\phi \rangle \right]\ ,
\label{fm21}
\end{eqnarray}
it satisfies the following equation:
\begin{eqnarray}
U'|\phi \rangle &=& {G_{0}}^{-1}P|\phi \rangle+P \widetilde{t}G_{0}U'|\phi \rangle \ .
\label{fm22}
\end{eqnarray}
We set $Z=\langle \phi|{G_{0}}^{-1}P|\phi \rangle $ and multiply \eq{fm22} by $P|\phi \rangle~Z^{-1}~\langle \phi|$
from the left-hand side (l.h.s.), and obtain
\begin{eqnarray}
P|\phi \rangle~Z^{-1}~\langle \phi|U'|\phi \rangle=P|\phi \rangle+P|\phi \rangle~Z^{-1}~\langle
\phi|P \widetilde{t}G_{0}U'|\phi \rangle \ .
\label{fm23}
\end{eqnarray}
We further multiply \eq{fm22} by $G_{0}$ from the l.h.s. and subtract \eq{fm23} from the result.
Then, the first terms of the r.h.s. cancel, and we obtain
\begin{eqnarray}
G_{0}U'|\phi \rangle &=& P|\phi \rangle~Z^{-1}~\langle \phi|U'|\phi \rangle+W \widetilde{t}G_{0}U'|\phi \rangle \ ,
\label{fm24}
\end{eqnarray}
where we have defined $W$ by
\begin{eqnarray}
W=G_{0}P-P|\phi \rangle~Z^{-1}~\langle \phi|P\ .
\label{fm25}
\end{eqnarray}
Thus, if we set
\begin{eqnarray}
G_{0}U'|\phi \rangle=Q|\phi \rangle~Z^{-1}~\langle \phi|U'|\phi \rangle \ ,
\label{fm26}
\end{eqnarray}
$Q|\phi \rangle $ satisfies
\begin{eqnarray}
Q|\phi \rangle &=& P|\phi \rangle+W \widetilde{t}Q|\phi \rangle \ .
\label{fm27}
\end{eqnarray}
We should note that $W$ in \eq{fm25} satisfies $ \langle \phi|{G_{0}}^{-1}W=0$ and $W{G_{0}}^{-1}|\phi \rangle=0$.
Thus, if we multiply \eq{fm27} by $ \langle \phi|{G_{0}}^{-1}$ from the l.h.s., we obtain
\begin{eqnarray}
\langle \phi|{G_{0}}^{-1}Q|\phi \rangle=\langle \phi|{G_{0}}^{-1} P|\phi \rangle=Z\ ,
\label{fm28}
\end{eqnarray}
which is consistent with the definition of $Q|\phi \rangle $ in \eq{fm26}.
Furthermore, $ \widetilde{t}_{\gamma}$ in \eq{fm27} can be restored to $t_{\gamma}$ for the deuteron channel
$ \gamma=\gamma _{d}$ owing to $W{G_{0}}^{-1}|\phi \rangle=0$, thus allowing us to write $ \widetilde{t}$ as $t$.

We should note that the channel wave function $|\phi \rangle $ is actually $|\phi;(\ell S_{c})J J_{z} \rangle $
in the channel-spin representation with at most three possible configurations, i.e.,
$(\ell S_{c})J=(J+\3H,\3H)J,~(J-\H,\H)J$, and $(J-\H,\3H)J$ for the parity $ \pi=(-1)^{J-\H}$,
and $(J-\3H,\3H)J,~(J+\H,\H)J$, and $(J+\H,\3H)J$ for $ \pi=(-1)^{J+\H}$. Namely, we should understand
\begin{eqnarray}
|\phi;(\ell S_{c})JJ_{z}\rangle=|q_{0},\psi _{d};(\ell S_{c})JJ_{z}\rangle
=\sum _{\lambda}\int^{\infty}_{0}p^{2}\,d\,p~|p,q_{0},\gamma _{d}\rangle~\psi _{\gamma _{d}}(p)\ ,
\label{fm29}
\end{eqnarray}
using $ \gamma _{d}$ defined in \eq{fm8}. The basic relationship in \eq{fm28} implies that
$Q|\phi \rangle $ is a modification of $P|\phi \rangle $ by the effect of the nonsingular
interaction $ \widetilde{t}$, and the real symmetric matrix $Z$ in \eq{fm28}
plays an essential role in the following discussion. In particular, $Z^{-1}$ is a nonsingular matrix
as seen in \eq{fm49}.\footnote{This can be proved if $g_{\gamma _{d},\gamma'_{d}}(q_{0},q_{0};x)$ in \eq{fm42}
is a positive- or negative-definite function, which is true at least for the dominant $S$-wave deuteron component.}
Thus, if we define
\begin{eqnarray}
\widetilde{Q}|\phi \rangle=Q|\phi \rangle~Z^{-1}\ \ ,\qquad \widetilde{P}|\phi \rangle=P|\phi \rangle~Z^{-1}\ ,
\label{fm30}
\end{eqnarray}
and multiply \eq{fm27} by $Z^{-1}$ from the r.h.s., we obtain our final equation
\begin{eqnarray}
\widetilde{Q}|\phi \rangle &=& \widetilde{P}|\phi \rangle+W \widetilde{t}\widetilde{Q}|\phi \rangle \ .
\label{fm31}
\end{eqnarray}

In order to derive the scattering amplitude, we multiply \eq{fm22} by $ \langle \phi|$ from the l.h.s. and
use \eq{fm26}. Then, we find
\begin{eqnarray}
\langle \phi|U'|\phi \rangle=Z+\langle \phi|P \widetilde{t}Q|\phi \rangle ~Z^{-1}~\langle \phi|U'|\phi \rangle \ .
\label{fm32}
\end{eqnarray}
Then, if we set
\begin{eqnarray}
\langle \phi|X|\phi \rangle=Z^{-1}~\langle \phi|P \widetilde{t}Q|\phi \rangle~Z^{-1}
=\langle \phi|\widetilde{P}\widetilde{t}\widetilde{Q}|\phi \rangle\ ,
\label{fm33}
\end{eqnarray}
we find 
\begin{eqnarray}
\langle \phi|U'|\phi \rangle=Z~\left[1+\langle \phi|X|\phi \rangle \langle \phi|U'|\phi \rangle \right]
=\left[Z^{-1}-\langle \phi|X|\phi \rangle \right]^{-1}\ .
\label{fm34}
\end{eqnarray}
This expression, together with \eq{fm21}, yields
\begin{eqnarray}
\langle \phi|U|\phi \rangle=\left[1+ic \langle \phi|U'|\phi \rangle \right]^{-1}\langle \phi|U'|\phi \rangle
=\left[Z^{-1}-\langle \phi|X|\phi \rangle+ic \cdot{\bf 1}\right]^{-1}\ .
\label{fm35}
\end{eqnarray}
A prescription of the principal-value integral still remains in \eq{fm33} for the $ \gamma=\gamma _{d}$ term.
The residue at $q=q_{0}$ is
\begin{eqnarray}
& & \sum _{\gamma _{d}}{q_{0}}^{2}\langle \phi|P|q_{0},\gamma _{d}\rangle
\widehat{t}_{\gamma _{d}}[q_{0}]\langle q_{0},\gamma _{d}|Q|\phi \rangle \nonumber \\
& & =\sum _{\gamma _{d},\gamma'_{d}}{q_{0}}^{2}\langle \phi|P|q_{0},\gamma _{d}\rangle
{g_{0}(\varepsilon _{d})}^{-1}|\psi _{\gamma _{d}}\rangle \langle \psi _{\gamma'_{d}}|
{g_{0}(\varepsilon _{d})}^{-1}\langle q_{0},\gamma'_{d}|Q|\phi \rangle \nonumber \\
& & ={q_{0}}^{2}\,\langle \phi|P{G_{0}}^{-1}|\phi \rangle \langle \phi|{G_{0}}^{-1}Q|\phi \rangle={q_{0}}^{2}Z^{2}\ ,
\label{fm36}
\end{eqnarray}
where \eq{fm28} is used in the last part. Thus, we only need to subtract \eq{fm36} since
$ \int^{\infty}_{0}d\,q~1/({q_{0}}^{2}-q^{2})=0$:
\begin{eqnarray}
& & \langle \phi|X|\phi \rangle=\sum _{\gamma \notin(\gamma _{d})}
\langle \phi|\widetilde{P}|\gamma \rangle~t_{\gamma}~\langle \gamma|\widetilde{Q}|\phi \rangle \nonumber \\
& & + \frac{4M}{3 \hbar^{2}}\int^{\infty}_{0}d\,q~\frac{1}{{q_{0}}^{2}-q^{2}}
\left[q^{2}\sum _{\gamma _{d}}\langle \phi|\widetilde{P}|q,\gamma _{d}\rangle~\widehat{t}_{\gamma _{d}}[q]
~\langle q,\gamma _{d}|\widetilde{Q}|\phi \rangle -{q_{0}}^{2}\right]\ .
\label{fm37}
\end{eqnarray}

We should note that the final equation (\ref{fm31}) still contains a delta function from the rearrangement factor
\begin{eqnarray}
\langle p,q,\gamma|P|p',q',\gamma'\rangle=\frac{1}{2}\int^{1}_{-1}dx~\frac{\delta(p-p_{1})}{p^{\lambda+2}}
g_{\gamma,\gamma'}(q,q',x)\frac{\delta(p'-p_{2})}{p'^{\lambda'+2}}\ .
\label{fm41}
\end{eqnarray}
Here, $p_{1}=p(q',q/2;x)$ and $p_{2}=p(q,q'/2;x)$ with $p(a,b;x)=\sqrt{a^{2}+b^{2}+2abx}$.
We use the spline interpolation for the variable $p$ and introduce the weight factors for the Gauss-Legendre quadrature.
For these, we use the notations
\begin{eqnarray}
& & p_{\mu \nu \kappa}=p(q_{\mu},q_{\nu}/2,x_{\kappa})
=\sqrt{{q_{\mu}}^{2}+{q_{\nu}}^{2}/4+q_{\mu}q_{\nu}x_{\kappa}}\ , \nonumber \\
& & S_{i \mu \nu \kappa}=S_{i}\left(p_{\mu \nu \kappa}\right)\ ,
\qquad (\hbox{coefficients~for~spline~interpolation}) \nonumber \\
& & B_{\gamma \mu, \gamma'\nu, \kappa}=q_{\mu}\sqrt{\omega _{\mu}}q_{\nu}\sqrt{\omega _{\nu}}
\left(\frac{1}{p_{\nu \mu \kappa}}\right)^{\lambda}\frac{1}{2}g_{\gamma,\gamma'}
\left(q_{\mu},q_{\nu};x_{\kappa}\right)\left(\frac{1}{p_{\mu \nu \kappa}}\right)^{\lambda'}\omega _{\kappa}\ ,
\label{fm42}
\end{eqnarray}
where $q_{\mu}$, $x_{\kappa}$, etc. are Gauss-Legendre discretization points and
$ \omega _{\mu}$, $ \omega _{\kappa}$, etc. are the corresponding weights.
The two-nucleon relative angular momenta, $ \lambda $ and $ \lambda'$, in $B_{\gamma \mu,\gamma'\nu,\kappa}$
are related to $ \gamma $ and $ \gamma'$, respectively.
For the on-shell momentum $q_{\mu}\rightarrow q_{0}$, we assume $q_{0}\sqrt{\omega _{0}}=1$.
Similarly, for $ \kappa=0$, appearing in the pole prescription for the $x$-integral later, we assume $ \omega _{0}=1$.
The weight factors are introduced, for example, as
\begin{eqnarray}
W_{i \mu \gamma,j \nu \gamma'}=q_{\mu}\sqrt{\omega _{\mu}}~{p_{i}}^{2}\omega _{i}
~\langle p_{i},q_{\mu},\gamma|W|p_{j},q_{\nu},\gamma'\rangle~{p_{j}}^{2}\omega _{j}~q_{\nu}\sqrt{\omega _{\nu}}\ .
\label{fm43}
\end{eqnarray}
The key relationship to avoid the appearance of the delta function for $p$ is the replacement
\begin{eqnarray}
\omega _{j}~\delta \left(p_{j}-p_{\mu \nu \kappa}\right)=S_{j \mu \nu \kappa}\ ,
\label{fm44}
\end{eqnarray}
which can be proved from the integration formula
\begin{eqnarray}
\int^{\infty}_{0}p'^{2}\,dp'~\frac{\delta(p'-p_{\mu \nu \kappa})}{p'^{2}}~f(p')=f(p_{\mu \nu \kappa})
=\sum _{j}S_{j \mu \nu \kappa}~f(p_{j})\ ,
\label{fm45}
\end{eqnarray}
for arbitrary smooth functions $f(p)$. For the spatial part of the deuteron wave functions, 
$ \widetilde{F}_{\lambda}(p)=\psi _{\gamma _{d}}(p)$, we also use the notation
$ \widetilde{F}_{j \lambda}=\widetilde{F}_{\lambda}(p_{})=F_{\lambda}(p_{j})/p_{j}$,
but apply the spline interpolation not to $ \widetilde{F}_{\lambda}(p)$ directly,
but to $ \langle p|{g_{0}}^{-1}(\varepsilon _{d})\psi _{d}\rangle $.
This is to avoid the numerical inaccuracy of the spline interpolation and guarantee the exact relationship
$W{G_{0}}^{-1}|\phi \rangle=P|\phi \rangle-P|\phi \rangle=0$. Namely, we define
\begin{eqnarray}
g_{j \lambda}=\langle p_{j}|{g_{0}}^{-1}(\varepsilon _{d})\psi _{d}\rangle
=\left(\varepsilon _{d}-\frac{\hbar^{2}}{M}p^{2}_{j}\right)\widetilde{F}_{j \lambda}\ ,
\label{fm46}
\end{eqnarray}
and calculate $ \widetilde{F}_{\lambda \mu 0 \kappa}$ through
\begin{eqnarray}
\widetilde{F}_{\lambda \mu 0 \kappa}=\left(\varepsilon _{d}-\frac{\hbar^{2}}{M}p^{2}_{\mu 0 \kappa}\right)^{-1}
\sum _{j}S_{j \mu 0 \kappa}~g_{j \lambda}\ ,
\label{fm47}
\end{eqnarray}
instead of $ \widetilde{F}_{\lambda \mu 0 \kappa}=\sum _{j}S_{j \mu 0 \kappa}~\widetilde{F}_{j \lambda}$.
Similarly, we calculate $ \widetilde{F}_{\lambda 0 \nu \kappa}$ through
\begin{eqnarray}
\widetilde{F}_{\lambda 0 \nu \kappa}=\left(\varepsilon _{d}-\frac{\hbar^{2}}{M}p^{2}_{0 \nu \kappa}\right)^{-1}
\sum _{j}S_{j 0 \nu \kappa}~g_{j \lambda}\ .
\label{fm48}
\end{eqnarray}
Then, by using the definition, $P_{i \mu \gamma}={p_{i}}^{2}\omega _{i}~q_{\mu}\sqrt{\omega _{\mu}}
~\langle p_{i},q_{\mu},\gamma|P|\phi \rangle $, we calculate $P_{i \mu \gamma}$ and $Z$ through
\begin{eqnarray}
& & P_{i \mu \gamma}=\sum _{\kappa}\sum _{\lambda'}S_{i 0 \mu \kappa}B_{\gamma \mu,\gamma'_{d}0,\kappa}
\widetilde{F}_{\lambda'\mu 0 \kappa}\ ,\nonumber \\
& & Z=\sum _{i \lambda}g_{i \lambda}P_{i0 \gamma _{d}}=\sum _{\kappa}\left(\varepsilon _{d}-\frac{\hbar^{2}}{M}
p^{2}_{00 \kappa}\right)\sum _{\lambda,\lambda'}\widetilde{F}_{\lambda 00 \kappa}~B_{\gamma _{d}0,\gamma'_{d}0, \kappa}
~\widetilde{F}_{\lambda'00 \kappa}\ .
\label{fm49}
\end{eqnarray}
For $ \widetilde{P}_{i \mu \gamma}=P_{i \mu \gamma}~Z^{-1}$, we trivially obtain
\begin{eqnarray}
U_{P}=\sum _{i \lambda}g_{i \lambda}\widetilde{P}_{i0 \gamma _{d}}=Z~Z^{-1}=1\ .
\label{fm50}
\end{eqnarray}
For the two-body $t$-matrix, we define
\begin{eqnarray}
t^{\gamma \widetilde{\gamma}}_{i j \mu}&=&\langle p_{i}|t^{It}_{(\lambda s),(\widetilde{\lambda}s)}
\left(\frac{3 \hbar^{2}}{4M}({q_{0}}^{2}-{q_{\mu}}^{2})+\varepsilon _{d}\right)|p_{j}\rangle \nonumber \\
&=& \frac{4 \pi}{(2 \pi)^{3}}~t^{It}_{(\lambda s),(\widetilde{\lambda}s)}
\left(p_{i},p_{j};\frac{3 \hbar^{2}}{4M}({q_{0}}^{2}-{q_{\mu}}^{2})+\varepsilon _{d}\right)\ ,
\label{fm51}
\end{eqnarray}
which is simply denoted by $t^{\gamma}_{i j \mu}$.

The discretization points for $q_{\mu}$ are divided into at least two or three regions,
depending on the incident energy. For $q_{0}<\kappa _{d}$ with $ \kappa _{d}=\sqrt{(4M/3 \hbar^{2})|\varepsilon _{d}|}$
(the neutron incident energy $E_{\rm cm}=(3 \hbar^{2}/4M){q_{0}}^{2}< |\varepsilon _{d}|$),
we only have the $nd$ elastic scattering, and the behavior of the two-body $t$-matrix is
\begin{eqnarray}
\begin{array}{cccc}
q_{\mu}: & [0,q_{0}] & q_{0} & [q_{0},\infty) \\ [2mm]
t:     & \hbox{real} \nearrow & \infty & \searrow \\
\end{array}\ .
\label{fm52}
\end{eqnarray}
For $q_{0}>\kappa _{d}$ ($E_{\rm cm}>|\varepsilon _{d}|$), the three-body breakup is possible
and the three-body Green function $G_{0}$ has a pole. The behavior of the two-body $t$-matrix in this case is
\begin{eqnarray}
\begin{array}{ccccc}
q_{\mu}: & [0,\sqrt{{q_{0}}^{2}-{\kappa _{d}}^{2}}] & [\sqrt{{q_{0}}^{2}-{\kappa _{d}}^{2}},q_{0}]
& q_{0} & [q_{0},\infty) \\ [2mm]
t:     & \hbox{complex} & \hbox{real} \nearrow & \infty & \searrow \\
\end{array}\ ,
\label{fm53}
\end{eqnarray}
so that $q_{M}=\sqrt{{q_{0}}^{2}-{\kappa _{d}}^{2}}$ is the threshold momentum for the deuteron breakup.
The three-body Green function pole is treated in accordance with the subtraction method 
discussed in Gl{\"o}ckle et al.'s review paper.\cite{PREP}

With these preparations, we can now write down the expressions for the numerical calculations.
The most important matrix elements are 
\begin{eqnarray}
& & A_{i \mu \gamma,j \nu \gamma'}=-q_{\mu}\sqrt{\omega _{\mu}}~{p_{i}}^{2}\omega _{i}
~\langle p_{i},q_{\mu},\gamma|PG_{0}|p_{j},q_{\nu},\gamma'\rangle~{p_{j}}^{2}\omega _{j}~q_{\nu}\sqrt{\omega _{\nu}}
=\frac{M}{\hbar^{2}q_{\mu}q_{\nu}} \nonumber \\
& & \times \left \{\begin{array}{lll}
\sum _{\kappa}\frac{1}{x_{\kappa}-x_{0 \mu \nu}}S_{i \nu \mu \kappa}B_{\gamma \mu,\gamma'\nu,\kappa}
S_{j \mu \nu \kappa} & \hbox{for} & |x_{0 \mu \nu}|>1\ ,\\ [2mm]
\sum _{\kappa}\frac{1}{x_{\kappa}-x_{0 \mu \nu}}\left[S_{i \nu \mu \kappa}B_{\gamma \mu,\gamma'\nu,\kappa}
S_{j \mu \nu \kappa}-S_{i \nu \mu 0}B_{\gamma \mu,\gamma'\nu,0}S_{j \mu \nu 0}\omega _{\kappa}\right] & & \\ [2mm]
+S_{i \nu \mu 0}B_{\gamma \mu,\gamma'\nu,0}S_{j \mu \nu 0}~\left[{\rm log}\left|\frac{1-x_{0 \mu \nu}}{1+x_{0 \mu \nu}}
\right|+i \pi \theta(1-|x_{0 \mu \nu}|) \right] & \hbox{for} & |x_{0 \mu \nu}|<1\ ,\\ [2mm]
\end{array} \right. \nonumber \\
\label{fm54}
\end{eqnarray}
with $x_{0 \mu \nu}=\left((3/4){q_{M}}^{2}-{q_{\mu}}^{2}-{q_{\nu}}^{2}\right)/(q_{\mu}q_{\nu})$.
The matrix elements of $W$ in \eq{fm43} are calculated from 
\begin{eqnarray}
W_{i \mu \gamma,j \nu \gamma'}=A_{i \mu \gamma,j \nu \gamma'}+\widetilde{P}_{i \mu \gamma}~P_{j \nu \gamma'}\ .
\label{fm55}
\end{eqnarray}
We can prove that
\begin{eqnarray}
\sum _{i \lambda}g_{i \lambda}~W_{i0 \gamma _{d},j \nu \gamma'}=0\ \ ,\qquad 
\sum _{j \lambda'} W_{i \mu \gamma,j0 \gamma'_{d}}~g_{j \lambda'}=0\ .
\label{fm56}
\end{eqnarray}
We also define the main kernel for the linear equation by
\begin{eqnarray}
M_{i \mu \gamma,j \nu \gamma'}=\sum _{k}W_{i \mu \gamma,k \nu \gamma'}~\widetilde{t}^{\gamma'}_{kj \nu}\ .
\label{fm57}
\end{eqnarray}
For the numerical check, we extend the definition of $M_{i \mu \gamma,j \nu \gamma'}$ in \eq{fm57} to include
\begin{eqnarray}
M_{i \mu \gamma,j0 \gamma _{d}}=\sum _{k}W_{i \mu \gamma,k0 \gamma _{d}}~\widehat{t}^{\gamma _{d}}_{kj0}=0\ ,
\label{fm58}
\end{eqnarray}
for the on-shell $ \gamma'=\gamma _{d}$ and $ \nu=0$, where the residue of the off-shell $t$-matrix
$ \widehat{t}^{\gamma _{d}}_{kj0}$ in the deuteron channel is given by (a separable kernel)
\begin{eqnarray}
\widehat{t}^{\gamma _{d}}_{kj0}=\widehat{t}^{\gamma _{d}\widetilde{\gamma}_{d}}_{kj0}
=g_{k \lambda}~g_{j \widetilde{\lambda}}\ .
\label{fm59}
\end{eqnarray}
The basic AGS equation \eq{fm31} is then expressed as
\begin{eqnarray}
\sum _{j \nu \gamma'} \left[\delta _{i,j}\delta _{\mu,\nu}\delta _{\gamma,\gamma'}
+M_{i \mu \gamma,j \nu \gamma'}\right]\widetilde{Q}_{j \nu \gamma'}=\widetilde{P}_{i \mu \gamma}\ .
\label{fm60}
\end{eqnarray}
For $ \widetilde{Q}_{i0 \gamma _{d}}$ calculated from 
\begin{eqnarray}
\widetilde{Q}_{i0 \gamma _{d}}=\widetilde{P}_{i0 \gamma _{d}}-\sum _{j \nu \gamma'} M_{i0 \gamma _{d},j \nu \gamma'}
\widetilde{Q}_{j \nu \gamma'}\ ,
\label{fm61}
\end{eqnarray}
we can show that 
\begin{eqnarray}
U_{Q}=\sum_{i \lambda}g_{i \lambda}~\widetilde{Q}_{i0 \gamma _{d}}=1\ ,
\label{fm62}
\end{eqnarray}
similarly to \eq{fm50}. The explicit expression of \eq{fm37} is 
\begin{eqnarray}
X_{\ell S_{c},\ell'S'_{c}} &=& \langle \phi _{\ell S_{c}}|X|\phi _{\ell'{S_{c}}'}\rangle
=\sum _{\gamma \notin (\gamma _{d})}\sum _{\mu,i,j}\widetilde{P}_{i \mu \gamma}
~t^{\gamma}_{ij \mu}~\widetilde{Q}_{j \mu \gamma}\nonumber \\
& & + \frac{4M}{3 \hbar^{2}}\sum _{\mu}\frac{1}{{q_{0}}^{2}-{q_{\mu}}^{2}}
\left[\sum _{\gamma _{d},i,j}\widetilde{P}_{i \mu{\gamma _{d}}}~\widehat{t}^{\gamma _{d}}_{ij \mu}
~\widetilde{Q}_{j \mu{\gamma _{d}}}-{q_{0}}^2 \omega _{\mu}\right]\ ,
\label{fm63}
\end{eqnarray}
and the $S$-matrix, $S^{J}_{(\ell'S'_{c}),(\ell S_c)}=\delta _{\ell,\ell'}\delta _{S_{c},S'_{c}}
-i2c~U^{J}_{(\ell'S'_{c}),(\ell S_{c})}$, is simply obtained by solving an equation
\begin{eqnarray}
\ \hspace{-10mm}& & \sum _{\ell',S_{c}'}\left[\left(Z^{-1}\right)_{\ell S_{c},\ell'S_{c}'}
-X_{\ell S_{c},\ell'S'_{c}}+ic~\delta _{\ell,\ell'}\delta _{S_{c},S_{c}'}\right]U^{J}_{(\ell'S_{c}'),
(\ell''S_{c}'')}=\delta _{\ell,\ell''}\delta _{S_{c},S_{c}''}\ .
\label{fm65}
\end{eqnarray}

\subsection{Optical theorem}
The original AGS equation contains the full information of the unitarity for the three-body scattering.
Here, we derive the optical theorem, starting from the AGS equation \eq{qm7} or
\begin{eqnarray}
U={G_{0}}^{-1}P+PtG_{0}U\ ,\label{op1}
\end{eqnarray}
for the systems composed of three identical particles. We first assume that there is no deuteron pole in the two-body
$t$-matrix $t$. We take the hermitian conjugate of \eq{op1} and replace ${G_{0}}^{-1}P$ with $U-PtG_{0}U$:
\begin{eqnarray}
U^{\dagger} &=& {G_{0}}^{-1}P+U^{\dagger}{G_{0}}^{\dagger}t^{\dagger}P
={G_{0}}^{-1}P+U^\dagger{G_{0}}^{\dagger}t^{\dagger}G_{0}\cdot{G_{0}}^{-1}P \nonumber \\
&=& {G_{0}}^{-1}P+U^{\dagger}{G_{0}}^{\dagger}t^{\dagger}G_{0}U-U^{\dagger}{G_{0}}^{\dagger}t^{\dagger}G_{0}PtG_{0}U\ .
\label{op2}
\end{eqnarray}
We again take the hermitian conjugate and use the fact that $P$ and $G_{0}$ are exchangeable,
\begin{eqnarray}
U &=& {G_{0}}^{-1}P+U^{\dagger}{G_{0}}^{\dagger}tG_{0}U-U^{\dagger}{G_{0}}^{\dagger}t^{\dagger}{G_{0}}^{\dagger}PtG_{0}U\ .
\label{op3}
\end{eqnarray}
The subtraction of \eq{op3} from \eq{op2} yields
\begin{eqnarray}
U^{\dagger}-U=U^{\dagger}{G_{0}}^{\dagger}(t^{\dagger}-t)G_{0}U
-U^{\dagger}{G_{0}}^{\dagger}t^{\dagger}(G_{0}-{G_{0}}^{\dagger})PtG_{0}U\ .
\label{op4}
\end{eqnarray}
Here, we use a basic relationship for the two-body $t$-matrix,
\begin{eqnarray}
t^{\dagger}-t=t^{\dagger}({G_{0}}^{\dagger}-G_{0})t
\label{op5}
\end{eqnarray}
and
\begin{eqnarray}
{G_{0}}^{\dagger}-G_{0}=2 \pi i \delta(E-H_{0})\ .
\label{op6}
\end{eqnarray}
Then, we find
\begin{eqnarray}
U^{\dagger}-U=2 \pi i~U^{\dagger}{G_{0}}^{\dagger}t^{\dagger}\delta(E-H_{0})(1+P)tG_{0}U\ .
\label{op7}
\end{eqnarray}

If we have a deuteron pole (or some equivalent divergence of the two-body $t$-matrix),
we have to separate this divergence term using \eq{fm18}.
Since $ \widetilde{t}$ does not involve the divergence, we can apply the above discussion
to $U'$ in \eq{fm22} and $ \widetilde{t}$, and obtain
\begin{eqnarray}
U'^{\dagger}-U'=2 \pi i~U'^{\dagger}{G_{0}}^{\dagger}\widetilde{t}^{\dagger}\delta(E-H_{0})(1+P)
\widetilde{t}G_{0}U'\ .
\label{op8}
\end{eqnarray}
If we multiply \eq{op8} by $[1+ic \langle \phi|U|\phi \rangle^{*}]\langle \phi|$ from the l.h.s.
and by $|\phi \rangle[1-ic \langle \phi|U|\phi \rangle]$ from the r.h.s., the relationship in \eq{fm21} yields
\begin{eqnarray}
\langle \phi|U|\phi \rangle^{*}-\langle \phi|U|\phi \rangle &=& 2i c \langle \phi|U|\phi \rangle^{*}
\langle \phi|U|\phi \rangle \nonumber \\
& & +2 \pi i~\langle \phi|U^{\dagger}{G_{0}}^{\dagger}\widetilde{t}^\dagger
\delta(E-H_0)(1+P)\widetilde{t}G_{0}U|\phi \rangle \ .
\label{op9}
\end{eqnarray}

We can derive another expression similar to \eq{op9}, starting from
\begin{eqnarray}
\left[Z^{-1}-\langle \phi|X|\phi \rangle+ic \right]~\langle \phi|U|\phi \rangle=1\ ,
\label{op10}
\end{eqnarray}
in \eq{fm35}. We multiply \eq{op10} by $ \langle \phi|U|\phi \rangle^{*}$ from the l.h.s.:
\begin{eqnarray}
\langle \phi|U|\phi \rangle^{*}=\langle \phi|U|\phi \rangle^{*}~\left[Z^{-1}-\langle \phi|X|\phi \rangle
+ic \right]~\langle \phi|U|\phi \rangle \ .
\label{op11}
\end{eqnarray}
We further take the hermitian conjugate of \eq{op11} and subtract it from \eq{op11}. Then, we find
\begin{eqnarray}
\langle \phi|U|\phi \rangle^{*}-\langle \phi|U|\phi \rangle &=& 2ic~\langle \phi|U|\phi \rangle^{*}
\langle \phi|U|\phi \rangle \nonumber \\
& & +\langle \phi|U|\phi \rangle^{*}~\left[\langle \phi|X|\phi \rangle^{*}-\langle \phi|X|\phi \rangle \right]
~\langle \phi|U|\phi \rangle \ .
\label{op12}
\end{eqnarray}
Thus, the second term of \eq{op12} corresponds to the second term of \eq{op9}.
This implies that the imaginary part of $ \langle \phi|X|\phi \rangle $ is related to the breakup cross sections.
We can also derive this from the basic equation in \eq{fm31}. We make $ \langle \phi|\widetilde{P}$ from \eq{fm31} and
use this in \eq{fm33}. Then, we find
\begin{eqnarray}
\langle \phi|X|\phi \rangle=\langle \phi|\widetilde{Q}^{\dagger}\widetilde{t}\widetilde{Q}|\phi \rangle
-\langle \phi|\widetilde{Q}^{\dagger}\widetilde{t}^{\dagger}W^{\dagger}\widetilde{t}\widetilde{Q}|\phi \rangle \ .
\label{op13}
\end{eqnarray}
We further take the hermitian conjugate of \eq{op13} and subtract \eq{op13} from the result:
\begin{eqnarray}
\langle \phi|X|\phi \rangle^{*}-\langle \phi|X|\phi \rangle=\langle \phi|\widetilde{Q}^{\dagger}
(\widetilde{t}^{\dagger}-\widetilde{t})\widetilde{Q}|\phi \rangle
-\langle \phi|\widetilde{Q}^{\dagger}\widetilde{t}^{\dagger}(W-W^{\dagger})
\widetilde{t}\widetilde{Q}|\phi \rangle \ .
\label{op14}
\end{eqnarray}
Then, by using \eq{op5} for $ \widetilde{t}$ and $W^{\dagger}-W=({G_{0}}^{\dagger}-G_{0})P$
derived from \eq{fm25}, we eventually obtain
\begin{eqnarray}
\langle \phi|X|\phi \rangle^{*}-\langle \phi|X|\phi \rangle
&=& \langle \phi|\widetilde{Q}^{\dagger}\widetilde{t}^{\dagger}({G_{0}}^{\dagger}-G_{0})(1+P)\widetilde{t}
\widetilde{Q}|\phi \rangle \nonumber \\
&=& 2 \pi i~\langle \phi|\widetilde{Q}^{\dagger}\widetilde{t}^{\dagger}\delta(E-H_{0})(1+P)\widetilde{t}
\widetilde{Q}|\phi \rangle \ .
\label{op15}
\end{eqnarray}
Note that the first term of \eq{op14} gives the direct term of the breakup process,
while the second term gives the exchange term. The last expression of \eq{op15} corresponds to the second term
of \eq{op9} through \eq{op12}, since we have
\begin{eqnarray}
\ \hspace{-10mm} G_{0}U|\phi \rangle=Q|\phi \rangle Z^{-1}\langle \phi|U|\phi \rangle
=\widetilde{Q}|\phi \rangle \langle \phi|U|\phi \rangle \equiv \widehat{Q}|\phi \rangle \ .
\label{op16}
\end{eqnarray}
After all, we have obtained the relationship
\begin{eqnarray}
(-2i)~{\rm Im} \langle \phi|U|\phi \rangle &=&  2ic~ \int d \widehat{\bq}_{0}~|\langle \bq_{0},\psi _{d}|U|\phi \rangle|^{2}
\nonumber \\
& & +2 \pi i~\langle \phi|\widehat{Q}^{\dagger}\widetilde{t}^{\dagger}
\delta(E-H_{0})(1+P)\widetilde{t}\widehat{Q}|\phi \rangle \ ,
\label{op17}
\end{eqnarray}
where the angular integral over $ \widehat{\bq}_{0}$ in the intermediate states is explicitly written.

We should note that $ \widetilde{t}$ in \eq{op17} can be safely replaced with $t$,
owing to the existence of the energy-conserving delta function of $ \delta (E-H_{0})$.
In fact, if we use Eqs.\,(\ref{fm18}) and (\ref{fm28}), we find
\begin{eqnarray}
tQ|\phi \rangle= \widetilde{t}Q|\phi \rangle-ic~{G_{0}}^{-1}|\phi \rangle Z\ .
\label{op18}
\end{eqnarray}
Here, the second term does not contribute to the last term of \eq{op17},
since $ \delta(E-H_{0}){G_{0}}^{-1}|\psi \rangle =0$.
Thus, $ \widetilde{t} \widehat{Q}|\phi \rangle $ in \eq{op17} can be replaced by $t \widehat{Q}|\phi \rangle
=tG_{0}U|\phi \rangle =T|\phi \rangle $, where $T=tG_{0}U$ is the three-body $t$-matrix.
If we further use $(1+P)^{2}=3(1+P)$ and the three-body breakup amplitude, $U_{0}|\phi \rangle=(1+P)T|\phi \rangle $,
\eq{op17} can be equivalently expressed as
\begin{eqnarray}
& & (-2i)~{\rm Im} \langle \phi|U|\phi \rangle=2ic~\int d \widehat{\bq}_{0}
~|\langle \bq_{0},\psi _{d}|U|\phi \rangle|^{2} \nonumber \\
& & +2 \pi i~\frac{1}{3} \int d\bp \int d\bq~\delta \left(E-\frac{\hbar^{2}}{M}\bp^{2}-\frac{3 \hbar^{2}}{4M}\bq^{2}\right)
~|\langle \bp,\bq|U_{0}|\phi \rangle|^{2}\ .
\label{op19}
\end{eqnarray}
In the channel-spin representation, we take the spin sum for the initial spin states $|S_{c} S_{cz} \rangle $ and obtain
\begin{eqnarray}
& & (-2i)~\sum _{S_{c},S_{cz}}~{\rm Im} \langle \phi;S_{c}S_{cz}|U|\phi;S_{c}S_{cz}\rangle \nonumber \\
& & = 2ic~\int d \widehat{\bq}_{0}~\sum _{S_{c},S_{cz}}~\sum _{S'_{c},S'_{cz}}
~|\langle \bq_{0},\psi _{d};S'_{c},S'_{cz}|U|\phi;S_{c}S_{cz}\rangle|^{2} \nonumber \\
& & +2 \pi i~\frac{1}{3}\int d\bp \int d\bq~\delta \left(E-\frac{\hbar^{2}}{M}\bp^{2}-\frac{3 \hbar^{2}}{4M}\bq^{2}\right)
\sum _{S_{c},S_{cz}}\sum _{S'_{c},S'_{cz}}|\langle \bp,\bq;S'_{c},S'_{cz}|U_{0}|\phi;S_{c}S_{cz}\rangle|^{2}\ . \nonumber \\
\label{op21}
\end{eqnarray}
We can carry out the partial-wave decomposition using
\begin{eqnarray}
& & \langle \phi_{\bq_{f}};S'_{c}S'_{cz}|U|\phi_{\bq_{i}};S_{c}S_{cz}\rangle
=\sum _{\ell'\ell JJ_{z}}U^{J}_{(\ell'S'_{c}),(\ell S_{c})} \nonumber \\
& & \times \sum _{m'}\langle \ell'm'S'_{c}S'_{cz}|JJ_{z}\rangle Y_{\ell'm'}(\widehat{\bq}_{f})
~\sum _{m}\langle \ell mS_{c}S_{cz}|JJ_{z}\rangle Y^{*}_{\ell m}(\widehat{\bq}_{i})\ ,
\label{op22}
\end{eqnarray}
with $|\bq_{f}|=|\bq_{i}|=q_{0}$. Since the operator $U$ is $J_{z}$-independent, we can prove
\begin{eqnarray}
\sum _{S_{c},S_{cz}}~\langle \phi;S_{c}S_{cz}|U|\phi;S_{c}S_{cz}\rangle
=\frac{1}{4 \pi}\sum _{(\ell S_{c})J}(2J+1)U^{J}_{(\ell S_{c}),(\ell S_{c})}\ .
\label{op23}
\end{eqnarray}
In the first term of the r.h.s. in \eq{op21}, the $ \widehat{\bq}_{0}$ integral allows us to take the sum 
over $S_{cz}$ and $J_{z}$ in a similar way. In the second term of the r.h.s. in \eq{op21},
we first integrate over $ \widehat{\bp}$ and $ \widehat{\bq}$, and change the $S'_{c},S'_{cz}$ sum to the $ \gamma $ sum:
\begin{eqnarray}
& & 2 \pi i~\frac{1}{3} \int^{\infty}_{0}p^{2}\,dp \int^{\infty}_{0}q^2\,dq
~\delta \left(E-\frac{\hbar^{2}}{M}p^{2}-\frac{3 \hbar^{2}}{4M}q^{2}\right) \nonumber \\
& & \times \sum _{S_c,S_{cz}}\sum _{\gamma}~\langle \phi _{\bq_{i}};S_{c}S_{cz}|{U_{0}}^{\dagger}|p,q,\gamma \rangle
\langle p,q,\gamma|U_{0}|\phi _{\bq_{i}};S_{c}S_{cz} \rangle \ .
\label{op24}
\end{eqnarray}
Here, the $p$-integral can be carried out from the $ \delta $ function. We use 
\begin{eqnarray}
E-\frac{\hbar^{2}}{M}p^{2}-\frac{3 \hbar^{2}}{4M}q^{2}=\frac{3 \hbar^{2}}{4M}({q_{M}}^{2}-q^{2})-\frac{\hbar^2}{M}p^{2}
=\frac{\hbar^{2}}{M}(p^{2}_{0}-p^{2})\ ,
\label{op25}
\end{eqnarray}
with $p_{0}=p_{0}(q)=\sqrt{(3/4)({q_{M}}^{2}-q^{2})}$, which implies that $q \leq q_{M}=\sqrt{{q_{0}}^{2}-{\kappa _{d}}^{2}}$.
(This corresponds to the change from $ \widetilde{t}$ to $t$ before.) Thus, we find
\begin{eqnarray}
& & \int^{\infty}_{0}p^{2}\,dp \int^{\infty}_{0}q^{2}\,dq
~\delta \left(E-\frac{\hbar^{2}}{M}p^{2}-\frac{3 \hbar^{2}}{4M}q^{2}\right) \nonumber \\
& & =\int^{q_{M}}_{0}q^{2}\,dq \int^{\infty}_{0}p^{2}\,dp~\delta \left(\frac{2 \hbar^{2}}{M}p_{0}(p_{0}-p)\right) \nonumber \\
& & =\frac{M}{2 \hbar^{2}}\int^{q_{M}}_{0}\,dq~q^{2}p_{0}\ .
\label{op26}
\end{eqnarray}
Thus, \eq{op24} becomes
\begin{eqnarray}
& & 2 \pi i~\frac{1}{3}~\frac{M}{2 \hbar^{2}}\int^{q_{M}}_{0}dq~q^{2}p_{0}
~\sum _{S_{c},S_{cz}}\sum _{\gamma}~\langle \phi _{\bq_{i}};S_{c}S_{cz}|{U_{0}}^{\dagger}|p_{0},q,\gamma \rangle \nonumber \\
& & \times \langle p_{0},q,\gamma|U_{0}|\phi _{\bq_{i}};S_{c}S_{cz} \rangle \ .
\label{op27}
\end{eqnarray}
The rest is the same as in deriving \eq{op23}. After all, the partial-wave decomposition of \eq{op21} is given by
\begin{eqnarray}
& & \frac{(-2i)}{4 \pi}~\sum _{(\ell S_{c})J}(2J+1)~{\rm Im}U^{J}_{(\ell S_{c}),(\ell S_{c})}
= i~\frac{q_{0}M}{3 \hbar^{2}}~\sum _{(\ell S_{c})(\ell'S'_{c})J}(2J+1)~|U^{J}_{(\ell'S'_{c}),(\ell S_{c})}|^{2} \nonumber \\
& & +\frac{1}{4 \pi}~2 \pi i~\frac{1}{3}~\frac{M}{2 \hbar^{2}}\int^{q_{M}}_{0}dq~q^{2}p_{0}
~\sum _{(\ell S_{c})J}(2J+1)\sum _{\gamma}|\langle p_{0},q,\gamma|U_{0}|q_{0},\psi _{d};(\ell S_{c})JJ_{z}\rangle|^{2}\ .
\nonumber \\
\label{op28}
\end{eqnarray}
We multiply \eq{op28} by the overall factor $(1/i)(3 \hbar^{2}/q_{0}M)$,
\begin{eqnarray}
& & (-2)\frac{1}{4 \pi}\frac{3 \hbar^{2}}{q_{0}M}~\sum _{(\ell S_{c})J}(2J+1)~{\rm Im}U^{J}_{(\ell S_{c}),(\ell S_{c})}
=\sum _{(\ell S_{c})(\ell'S'_{c})J}(2J+1)~|U^{J}_{(\ell'S'_{c}),(\ell S_{c})}|^{2} \nonumber \\
& & +\frac{1}{4 \pi}\frac{\pi}{q_{0}}~\int^{q_{M}}_{0}dq~q^{2}p_{0}~\sum _{(\ell S_{c})J}(2J+1)
\sum _{\gamma}|\langle p_{0},q,\gamma|U_{0}|q_{0},\psi _{d};(\ell S_{c})JJ_{z}\rangle|^{2}\ ,
\label{op29}
\end{eqnarray}
and use the elastic and breakup scattering amplitudes with a common factor
\begin{eqnarray}
& & f^{J}_{(\ell'S'_{c}),(\ell S_{c})}=-\frac{1}{(4 \pi)}\frac{4M}{3 \hbar^{2}}\frac{(2 \pi)^{3}}{(4 \pi)}
U^{J}_{(\ell'S'_{c}),(\ell S_{c})}=-\frac{\pi}{2}\frac{4M}{3 \hbar^{2}}U^{J}_{(\ell'S'_{c}),(\ell S_{c})}\ ,\nonumber \\
& & f^{({\rm br})J}_{\gamma(\ell S_{c})}(q)=-\frac{\pi}{2}\frac{4M}{3 \hbar^{2}}\langle p_{0},q,\gamma|U_{0}
|q_{0},\psi _{d};(\ell S_{c})JJ_{z}\rangle \ .
\label{op30}
\end{eqnarray}
Multiplying \eq{op29} by the overall factor $4 \pi \{-(\pi/2)(4M/3 \hbar^{2})\}^{2}$, we find
\begin{eqnarray}
& & \frac{4 \pi}{q_{0}}\sum _{(\ell S_{c})J}(2J+1)~{\rm Im}f^{J}_{(\ell S_{c}),(\ell S_{c})}
=4 \pi \sum _{(\ell S_{c})(\ell'S'_{c})J}(2J+1)~|f^{J}_{(\ell'S'_{c}),(\ell S_{c})}|^{2} \nonumber \\
& & +\frac{\pi}{q_{0}}~\int^{q_{M}}_{0}dq~q^{2}p_0 ~\sum _{(\ell S_{c})J}(2J+1)\sum _{\gamma}
~|f^{({\rm br})J}_{\gamma(\ell S_{c})}(q)|^{2}\ .
\label{op31}
\end{eqnarray}
We find that the following relationship holds for each $(\ell S_{c})J$ component:
\begin{eqnarray}
\ \hspace{-10mm} \frac{1}{q_{0}}~{\rm Im}f^{J}_{(\ell S_{c}),(\ell S_{c})}=\sum _{(\ell'S'_{c})}
~|f^{J}_{(\ell'S'_{c}),(\ell S_{c})}|^{2}+\frac{1}{4q_{0}}\int^{q_{M}}_{0}dq~q^{2}p_{0} 
\sum _{\gamma}|f^{({\rm br})J}_{\gamma(\ell S_{c})}(q)|^{2}\ .\nonumber \\
\label{op32}
\end{eqnarray}
The relationship of the total cross sections is obtained by taking a spin average over the initial spin multiplicities
($(2I+1)(2s+1)=6$ with $I=1$ and $s=\H$), namely, by taking the sum $(4 \pi/6)\sum _{(\ell S_{c})J}(2J+1)$ over \eq{op32}.
For the practical calculation of the total breakup cross sections in \eq{op32}, it is most convenient to use
the imaginary part of $ \langle \phi|X|\phi \rangle $ through \eq{op15}. It is rather straightforward to derive
\begin{eqnarray}
\ \hspace{-10mm} & & \frac{1}{q_{0}}~{\rm Im}f^{J}_{(\ell S_{c}),(\ell S_{c})}
=\sum _{(\ell'S'_{c})}~|f^{J}_{(\ell'S'_{c}),(\ell S_{c})}|^{2} \nonumber \\
\ \hspace{-10mm} & & +\left(-\frac{2}{\pi}\right)\frac{\hbar^{2}}{M}\frac{3}{4q_{0}}
\sum _{(\ell'S'_{c}),(\ell''S''_{c})}{f^{J}_{(\ell'S'_{c}),(\ell S_{c})}}^{*}
~\hbox{Im}X^{J}_{(\ell'S'_{c}),(\ell''S''_{c})}~f^{J}_{(\ell''S''_{c}),(\ell S_{c})}\ ,\hfill
\label{op33}
\end{eqnarray}
where $X^{J}_{(\ell'S'_{c}),(\ell''S''_{c})}$ is given in \eq{fm63}.

\subsection{Moving singularity of the three-body Green function}
The direct solution of \eq{fm60} in terms of \eq{fm54} causes a serious numerical problem,
when it is applied to the energies above the three-body breakup threshold, namely,
$E_{\rm cm}=(3 \hbar^{2}/4M){q_{0}}^{2}>|\varepsilon _{d}|$.
If we use a restricted number of discretization points, say, 5 points for each interval of \eq{fm53},
it gives reasonable results. However, if we increase the number of discretization points,
the phase shift results of $^{4}S_{\3H}$ states, for example, deviate by more than several degrees.
The origin of this pathological situation is the discretization points close to the logarithmic singularity
in the kernel \eq{fm54}. This is a notorious problem of moving singularities,
appearing in any type of three-body model, that takes account of breakup processes.
We avoid this by using the prescription given by the Bochum-Krakow group,\cite{spline82,Wi03,PREP,Liu05} namely,
applying the spline interpolation even to the $q$ degree of freedom.
The main idea is that by applying the spline interpolation to the logarithmic and step function terms in \eq{fm54},
we can avoid the situation wherein the discretization points directly hit the boundary of the crescent-shape region
of the $q$-$q'$ plane. 

A general prescription is given in \S 4 of Liu et al.'s paper.\cite{Liu05}
We first separate the $q$-$q'$ plane into two regions, one is the rectangular region with $q<q_{M}$ and $q'<q_{M}$,
and the other is the region with $q>q_{M}$ or $q'>q_{M}$. Here, $q_{M}=\sqrt{{q_{0}}^{2}-{\kappa _{d}}^{2}}$ (see \eq{fm53}).
In the latter region, we can prove that $x_{0}$ defined by $x_{0}=x_{0}(q,q')=((3/4){q_{M}}^{2}-q^{2}-{q'}^{2})/(qq')$
is always less than $-1$, so that the unsubtracted expression in \eq{fm54} is safely used.
In the rectangular region, the subtraction is made in the following scheme.
First, let us consider the angular integral
\begin{eqnarray}
I(q,q')=\int^{1}_{-1}dx~\frac{F(q,q';x)}{x-x_{0}-i0}\ ,
\label{mv1}
\end{eqnarray}
for $q,~q'\leq q_{M}$. The kernel function $F(q,q';x)$ is composed of the $q,~q'$ dependence
from the spline interpolation for $p$ and $p'$, $G(q,q';x)$, and the rearrangement coefficients $B(q,~q';~x)$:
\begin{eqnarray}
F(q,q';x)=G(q,q';x)~B(q,q';x)\ . \label{mv2}
\end{eqnarray}
More explicitly, we move $(1/p_{\nu \mu \kappa})^{\lambda}$ and $(1/p_{\mu \nu \kappa})^{\lambda'}$ factors
in \eq{fm42} to $G(q,q';x)$, and assign
\begin{eqnarray}
& & G_{i,j}(q_{\mu},q_{\nu};x_{\kappa})=S_{i \nu \mu \kappa}\frac{1}{\left(p_{\nu \mu \kappa}\right)^{\lambda}}
~S_{j \mu \nu \kappa}\frac{1}{\left(p_{\mu \nu \kappa}\right)^{\lambda'}}\ ,\nonumber \\
& & B_{\gamma,\gamma'}(q_{\mu},q_{\nu};x_{\kappa})=q_{\mu}q_{\nu}\frac{1}{2}g_{\gamma,\gamma'}(q_{\mu},q_{\nu};x_{\kappa})\ .
\label{mv3}
\end{eqnarray}
The function $B(q,q';x)$ is expressed as a finite angular-momentum sum
\begin{eqnarray}
B(q,q';x)=\sum^{k_{M}}_{k=0}(2k+1)B^{(k)}(q,q')~P_{k}(x)\ ,
\label{mv4}
\end{eqnarray}
with $B^{(k)}(q,q')$ being a polynomial of $q$ and $q'$. We modify \eq{mv1} to
\begin{eqnarray}
I(q,q') &=& \int^{1}_{-1}dx~\frac{G(q,q';x)-G(q,q';x_{0})}{x-x_{0}}~B(q,q';x)\nonumber \\
& & +G(q,q';x_{0})\int^{1}_{-1}dx~\frac{B(q,q';x)}{x-x_{0}-i0}\ .
\label{mv5}
\end{eqnarray}
Then, the second integral is expressed using the Legendre function of the second kind,
$Q_{k}(z)\equiv(1/2)\int^{1}_{-1}dx~P_{k}(x)/(z-x)$. Thus, we find
\begin{eqnarray}
I(q,q') &=& \int^{1}_{-1} dx~\frac{G(q,q';x)-G(q,q';x_{0})}{x-x_{0}}~B(q,q';x)\nonumber \\
& & +(-2)~G(q,q';x_{0})\sum^{k_{M}}_{k=0}(2k+1)B^{(k)}(q,q')~Q_{k}(x_{0}+i0)\ ,
\label{mv6}
\end{eqnarray}
where
\begin{eqnarray}
& & Q_{k}(x_{0}+i0)=\left(-\frac{1}{2}\right)\left \{P_{k}(x_{0})\left[{\rm log}\left|\frac{1-x_{0}}{1+x_{0}}\right|
+i \pi \theta(1-|x_{0}|)\right]+2W_{k-1}(x_{0})\right \}\ . \nonumber \\
\label{mv7}
\end{eqnarray}
We are interested in the integral
\begin{eqnarray}
I=\int^{q_{M}}_{0}dq~\int^{q_{M}}_{0}dq'f(q)~I(q,q')~g(q')\ ,
\label{mv8}
\end{eqnarray}
with the spline interpolation
\begin{eqnarray}
f(q)=\sum _{\mu}S_{\mu}(q)f(q_{\mu}) \qquad \hbox{for} \quad q \leq q_{M} \qquad \hbox{etc.}\ ,
\label{mv9}
\end{eqnarray}
i.e., $I=\sum _{\mu,\nu}f_{\mu}~I_{\mu,\nu}~g_{\nu}$ with $f_{\mu}=\sqrt{\omega _{\mu}}~f(q_{\mu})$ etc.
For the integrals to calculate $I_{\mu,\nu}$, we can replace the integral variables $q$ and $q'$ with
$q_{\mu}$ and $q_{\nu}$ safely if the variation of the functions is sufficiently smooth in the mesh-point intervals. 
Thus, we find
\begin{eqnarray}
I_{\mu,\nu} &=& \frac{1}{\sqrt{\omega _{\mu}}\sqrt{\omega _{\nu}}}
\int^{q_{M}}_{0}dq~\int^{q_{M}}_{0}dq'~S_{\mu}(q)~I(q,q')~S_{\nu}(q')\nonumber \\
&=& \sqrt{\omega _{\mu}}\sqrt{\omega _{\nu}}\int^{1}_{-1}dx~\frac{G(q_{\mu},q_{\nu};x)-G(q_{\mu},q_{\nu};x_{0 \mu \nu})}
{x-x_{0 \mu \nu}}~B(q_{\mu},q_{\nu};x)\nonumber \\
& & +\sqrt{\omega _{\mu}}\sqrt{\omega _{\nu}}~G(q_{\mu},q_{\nu};x_{0 \mu \nu})
\sum^{k_{M}}_{k=0}(2k+1) B^{(k)}(q_{\mu},q_{\nu})~\widetilde{Q}_{k \mu \nu}\ ,
\label{mv10}
\end{eqnarray}
where
\begin{eqnarray}
\ \hspace{-10mm} \widetilde{Q}_{k \mu \nu}=(-2)\frac{1}{\omega _{\mu}\omega _{\nu}}
\int^{q_{M}}_{0}dq~\int^{q_{M}}_{0}dq'~S_{\mu}(q)Q_{k}(x_{0}+i0)S_{\nu}(q')\ .
\label{mv11}
\end{eqnarray}
Using these results, we replace the subtracted expression of $A_{i \mu \gamma,  \nu \gamma'}$ in \eq{fm54} with
\begin{eqnarray}
A_{i \mu \gamma,j \nu \gamma'} &=& \sum _{\kappa}\frac{1}{x_{\kappa}-x_{0 \mu \nu}}
\left[S_{i \nu \mu \kappa}S_{j \mu \nu \kappa}-S_{i \nu \mu 0}S_{j \mu \nu 0}
\left(\frac{p_{\nu \mu \kappa}}{p_{\nu \mu 0}}\right)^{\lambda}\left(\frac{p_{\mu \nu \kappa}}{p_{\mu \nu 0}}\right)^{\lambda'}
\right]~B_{\gamma \mu,\gamma' \nu,\kappa} \nonumber \\
& & +S_{i \nu \mu 0}S_{j \mu \nu 0}~\widetilde{B}_{\gamma \mu,\gamma'\nu}\ ,
\label{mv12}
\end{eqnarray}
where $ \widetilde{B}_{\gamma \mu,\gamma'\nu}$ is generated from $B_{\gamma \mu,\gamma'\nu,0}$
by just modifying $P_{k}(x_{0 \mu \nu})$ in $g_{\gamma,\gamma'}(q_{\mu},q_{\nu}; \linebreak x_{0 \mu \nu})$
to $ \widetilde{Q}_{k \mu \nu}$ in \eq{mv11}. Unfortunately, a complete analytical evaluation of \eq{mv11}
is not possible. Here, we use the one-side spline interpolation formula and give in Appendix B a detailed procedure to calculate
\begin{eqnarray}
\ \hspace{-10mm} Q_{k \mu \nu}=(-2)\frac{1}{\omega _{\nu}}~\int^{q_{M}}_{0}dq'
~Q_{k}(x_{0 \mu}+i0)S_{\nu}(q')\ ,
\label{mv13}
\end{eqnarray}
with $x_{0 \mu}=((3/4){q_{M}}^{2}-{q_{\mu}}^{2}-q'^{2})/(q_{\mu}q')$. Actually,
this procedure breaks the symmetry of $ \widetilde{Q}_{k \mu \nu}$ with respect to the exchange of $ \mu $ and $ \nu $.
This generic inaccuracy of the spline interpolation technique is, however, very small and
we recover the symmetry of the $S$-matrix at the stage of calculating $X_{\ell S_{c},\ell'S'_{c}}$ in \eq{fm63}, i.e.,
by modifying $X_{\ell S_{c},\ell'S'_{c}}$ to $(1/2)[X_{\ell S_{c},\ell'S'_{c}}+X_{\ell'S'_{c},\ell S_{c}}]$.\footnote{
Various methods to symmetrize $Q_{k \mu \nu}$ cause a serious numerical problem 
at the point $q=(\sqrt{3}/2)q_{M}$ for the solution of \eq{fm60}.} 

\section{Results and discussion}
\subsection{Total cross sections}
For the energies above the deuteron breakup threshold, we further separate the momentum intervals over $q$
in \eq{fm53} into the following 6 intervals:
\begin{eqnarray}
\begin{array}{ccccccc}
q_{\mu}: & \left[0,\frac{1}{2}q_{M}\right] & \left[\frac{1}{2}q_{M},\frac{\sqrt{3}}{2}q_{M}\right]
& \left[\frac{\sqrt{3}}{2}q_{M},q_{M}\right] & [q_{M},q_{0}] & [q_{0},2q_{0}] & [2q_{0},\infty) \\
\end{array}\ ,\nonumber \\
\label{to1}
\end{eqnarray}
where $q_{M}=\sqrt{{q_{0}}^{2}-{\kappa _{d}}^{2}}$. This is chosen from the criterion that
the $q_{\mu}$ points do not hit the positions of the logarithmic singularities and that
sufficient points cover the rapidly changing region of the kernel in \eq{mv12}.\footnote{
If we choose the interval $[0,q_{M}]$ with the odd number of Gauss-Legendre quadrature mesh points,
the middle point hits the logarithmic singularity point $(q_{M}/2,q_{M}/2)$. See \eq{b7}.}
The first three intervals are discretized with the $n_{1}$-point Gauss-Legendre quadrature,
and the next two intervals with the $n_{2}$-point Gauss-Legendre quadrature.
For the outermost interval $[2q_{0},\infty)$, we apply a mapping $q_{\mu}=2q_{0}+\hbox{tan}\left \{(\pi/4)(1+x_{i})\right \}$
with $x_{i} \in [-1,1]$ ($i=1$ - $n_{3}$) being the $n_{3}$-point Gauss-Legendre quadrature.
We have altogether $3n_{1}+2n_{2}+n_{3}$ mesh points for the whole $q$.
For the energies below the deuteron breakup threshold, we actually use
\begin{eqnarray}
\begin{array}{ccccccc}
q_{\mu}: & \left[0,\frac{1}{2}q_{0}\right] & \left[\frac{1}{2}q_{0},q_{0}\right] & [q_{0},2q_{0}] & [2q_{0},3] & [3,6]
& [6,\infty)
\end{array}\ ,
\label{to2}
\end{eqnarray}
where the unit is in $ \hbox{fm}^{-1}$ and $q_{0}$ is the incident momentum of the neutron
with $E=(3 \hbar^{2}/4M){q_{0}}^{2}+\varepsilon _{d}$. Each interval is discretized with the
$n_{1}$-$n_{2}$-$n_{3}$ point Gauss-Legendre quadrature in a similar way.
Since the maximum value of $q_{0}$ is 0.267~$ \hbox{fm}^{-1}$, $2q_{0}$ does not reach 3 $\hbox{fm}^{-1}$.
For the bound-state problem of negative energies, we use
\begin{eqnarray}
\begin{array}{ccccccc}
q_{\mu}: & [0,0.2] & [0.2,0.5] & [0.5,1] & [1,3] & [3,6] & [6,\infty) \\
\end{array}\ ,
\label{to3}
\end{eqnarray}
although such structure in the small momentum region may not be necessary.
For the $p$ mesh points, the previous four-interval separation for the bound state problem \cite{ren}
\begin{eqnarray}
\begin{array}{ccccc}
p_{i}: & [0,1] & [1,3] & [3,6] & [6,\infty) \\
\end{array}\ ,
\label{to4}
\end{eqnarray}
is used with the $n_{2}$-point (for the first three intervals) and
$n_{3}$-point (for the outermost interval) Gauss-Legendre quadratures.
Actual Faddeev calculations, however, are carried out using the discretization points only up to 6~$ \hbox{fm}^{-1}$,
to avoid the inaccuracy caused by the spline interpolation.
The high-momentum $p$ is, however, necessary for the accurate evaluation of the off-shell $t$-matrix,
so that another set of discretization points with 10+15 points is employed to calculate the $t$-matrix for the positive energies.
The middle point is the on-shell momentum $p_{0}=\sqrt{(3/4)({q_{M}}^{2}-{q_{0}}^{2})}$ in \eq{fm51},
to which the Noyes-Kowalski formalism is again conveniently applied.
For the negative energies, the discretization points in \eq{to4} are directly used.

We also need to consider the discretization points for the angular-momentum projection of the rearrangement coefficients
in Eqs.\,(\ref{fm54}) and (\ref{mv12}). For energies below the breakup threshold,
the previous 20-point Gauss-Legendre quadrature formula \cite{ren} is safely used for the Legendre polynomials,
since there is no singularity point for the $x$ integral.
For the energies above the breakup threshold, the $[-1,1]$ interval for $x$ is separated into two parts,
$[-1,x_{0 \mu \nu}]$ and $[x_{0 \mu \nu},1]$. For the larger interval, the 15-point Gauss-Legendre formula is used,
and for the smaller interval, the 5-point formula is used.

The three-body model space is mainly specified by the maximum value of the two-nucleon angular momentum
$I_{\rm max}$ in \eq{fm7}. The maximum orbital angular momentum for the two-nucleon subsystem
is therefore $ \lambda _{\rm max}=I_{\rm max}+1$.
We need to take a sufficient number of the relative angular momentum $ \ell $ corresponding to $ \widehat{\bq}$,
to reproduce the backward rise of the differential cross sections. Here, we assume
$ \ell _{\rm max}=\hbox{Min}\{2 \lambda _{\rm max},10\}$, which leads to the total angular momentum up to
$J_{\rm max}=\ell _{\rm max}+(S_{c})_{\rm max} \leq 3 \lambda _{\rm max}-\H$ since $(S_{c})_{\rm max}
=I_{\rm max}+\H=\lambda _{\rm max}-\H$. For the deuteron channels, however,
the maximum $J$ value is much smaller and $J_{\rm max}=\ell _{\rm max}+\3H \leq 2 \lambda _{\rm max}+\3H=2I_{\rm max}
+{\scriptstyle \frac{7}{2}}$. We also examine the model space composed of $^{3}S_{1}+\hbox{}^{3}D_{1}$
plus $^{1}S_{0}$ $NN$ interactions only, which corresponds to the so-called five-channel calculation
of the $J^\pi=\H^+$ bound state. We call this the $S+D$ model space.
For example, the three-body angular-momentum states in the deuteron channels are restricted up to
$^{4}G_{\scriptstyle \frac{11}{2}}$, $^{4}I_{\scriptstyle \frac{15}{2}}$, $^{4}K_{\scriptstyle \frac{19}{2}}$,
and $^{4}M_{\scriptstyle \frac{23}{2}}$, for $I_{\rm max}=S+D$, 2, 3, and 4, respectively, in the usual spectroscopic notation.

\begin{figure}[b]
\centerline{\includegraphics[width=0.5\columnwidth]{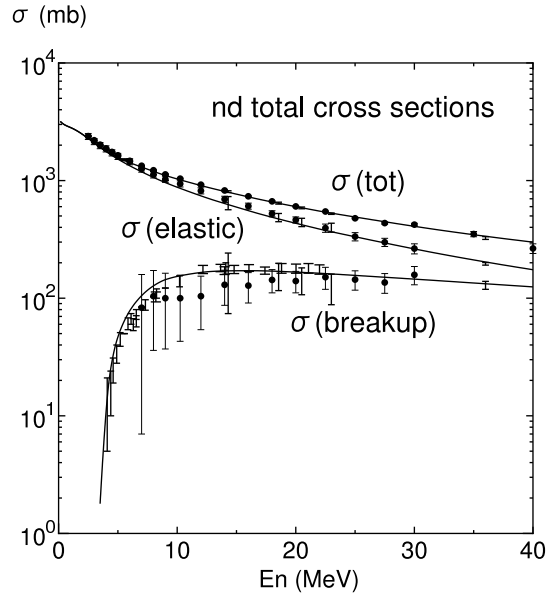}}
\caption{fss2 predictions to the $nd$ total cross sections up to $E_{n}=40$ MeV, compared with the experiment.
The experimental data were taken from Ref.~\citen{Sc83} for the filled circles with error bars
and Refs.~\citen{Ca61,Ho69,Pa75,Se72} for the others.}
\label{fig1}
\end{figure}

The elastic and breakup total cross sections up to $E_{n}=40$ MeV, predicted using fss2, are plotted in Fig.\,\ref{fig1},
together with the experimental data. Here, the incident neutron energy, $E_{n}=(3/2)E_{\rm cm}$,
is measured in the laboratory system. In this calculation, we have used $n \equiv n_{1}\hbox{-}n_{2}\hbox{-}n_{3}
=5 \hbox{-}6 \hbox{-}5$ and $I_{\rm max}=4$. Although some discrepancy might exist around $E_{n}=10$ MeV,\cite{PREP}
the elastic and breakup total cross sections are reasonably reproduced with the constraint of the optical theorem.

\subsection{Differential cross sections}
The differential cross sections for the $nd$ elastic scattering are calculated from the scattering amplitudes
$f^{J}_{(\ell'S'_{c}),(\ell S_{c})}$ in \eq{op30} by summing up over the final spin states and
by averaging over the initial spin states:
\begin{eqnarray}
& & \frac{d \sigma}{d \Omega}=\frac{1}{6}\sum _{S'_{c},S_{c}}\sum _{L}B_{L}(S'_{c},S_{c})P_{L}(\cos{\theta})\ ,\nonumber \\
& & B_{L}(S'_{c},S_{c})=(-1)^{S'_{c}-S_{c}}\sum _{J_{1}J_{2}\ell _{1}\ell _{2}\ell'_{1}\ell'_{2}}
Z(\ell'_{1}J_{1}\ell'_{2}J_{2};S'_{c}L)Z(\ell _{1}J_{1}\ell _{2}J_{2};S_{c}L)\nonumber \\
& & \ \hspace{2cm} \times{f^{J_{1}}_{(\ell'_{1}S'_{c}),(\ell _{1}S_{c})}}^{*}f^{J_{2}}_{(\ell'_{2}S'_{c}),(\ell _{2}S_{c})}\ ,
\label{dif1}
\end{eqnarray}
where the rearrangement factor is given by the Wigner-Racah coefficients as \cite{LT58}
\begin{eqnarray}
Z(\ell _{1}J_{1}\ell _{2}J_{2};S_{c}L)=\widehat{\ell}_{1}\widehat{\ell}_{2}\widehat{J}_{1}\widehat{J}_{2}
~\langle \ell _{1}0 \ell _{2}0|L0 \rangle~W(\ell _{1}J_{1}\ell _{2}J_{2};S_{c}L)\ .
\label{dif2}
\end{eqnarray}
The elastic total cross sections are obtained from $L=0$ components in \eq{dif1} by using a special case
$B_{0}(S'_{c},S_{c})=\sum _{J \ell \ell'}(2J+1)\vert f^{J}_{(\ell'S'_{c}),(\ell S_{c})}\vert^{2}$. We find
\begin{eqnarray}
\sigma _{e \ell}=\frac{4 \pi}{6}\sum _{S'_{c},S_{c}}B_{0}(S'_{c},S_{c})
=\frac{4 \pi}{6}\sum _{(\ell S_{c})J}(2J+1)\sum _{(\ell'S'_{c})}\vert f^{J}_{(\ell'S'_{c}),(\ell S_{c})}\vert^{2}\ ,
\label{dif3}
\end{eqnarray}
which is of course consistent with \eq{op32}.

\begin{figure}[htb]
\centerline{\includegraphics[width=0.98\textwidth]{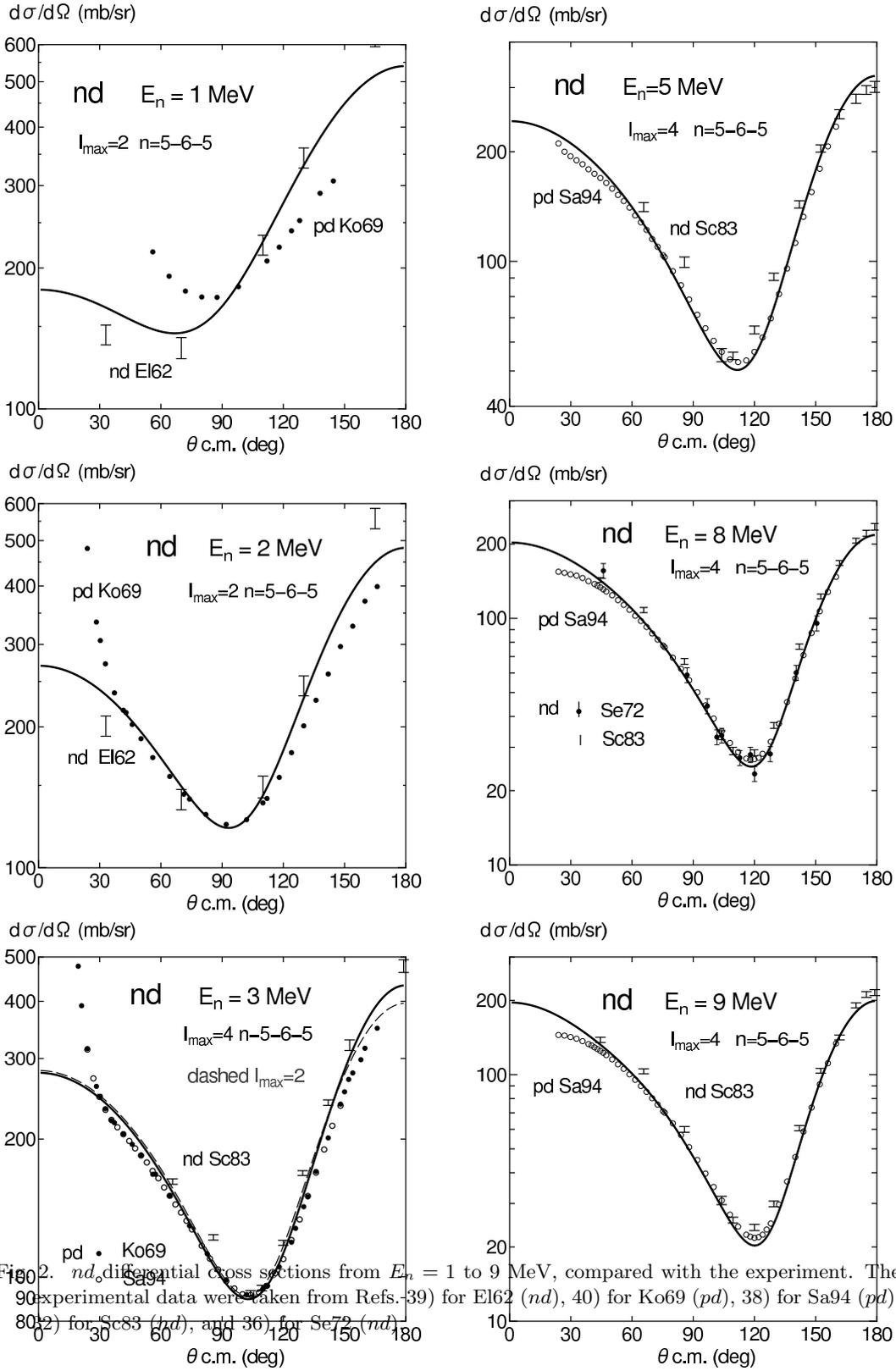}}
\vspace{-20mm}
\caption{$nd$ differential cross sections from $E_{n}=1$ to 9 MeV, compared with the experiment.
The experimental data were taken from Refs.~\citen{El62} for El62 ($nd$),
\citen{Ko69} for Ko69 ($pd$), \citen{Sa94} for Sa94 ($pd$), \citen{Sc83} for Sc83 ($nd$), and \citen{Se72} for Se72 ($nd$).}
\label{fig2}
\end{figure}

\begin{figure}[htb]
\centerline{\includegraphics[width=0.98\textwidth]{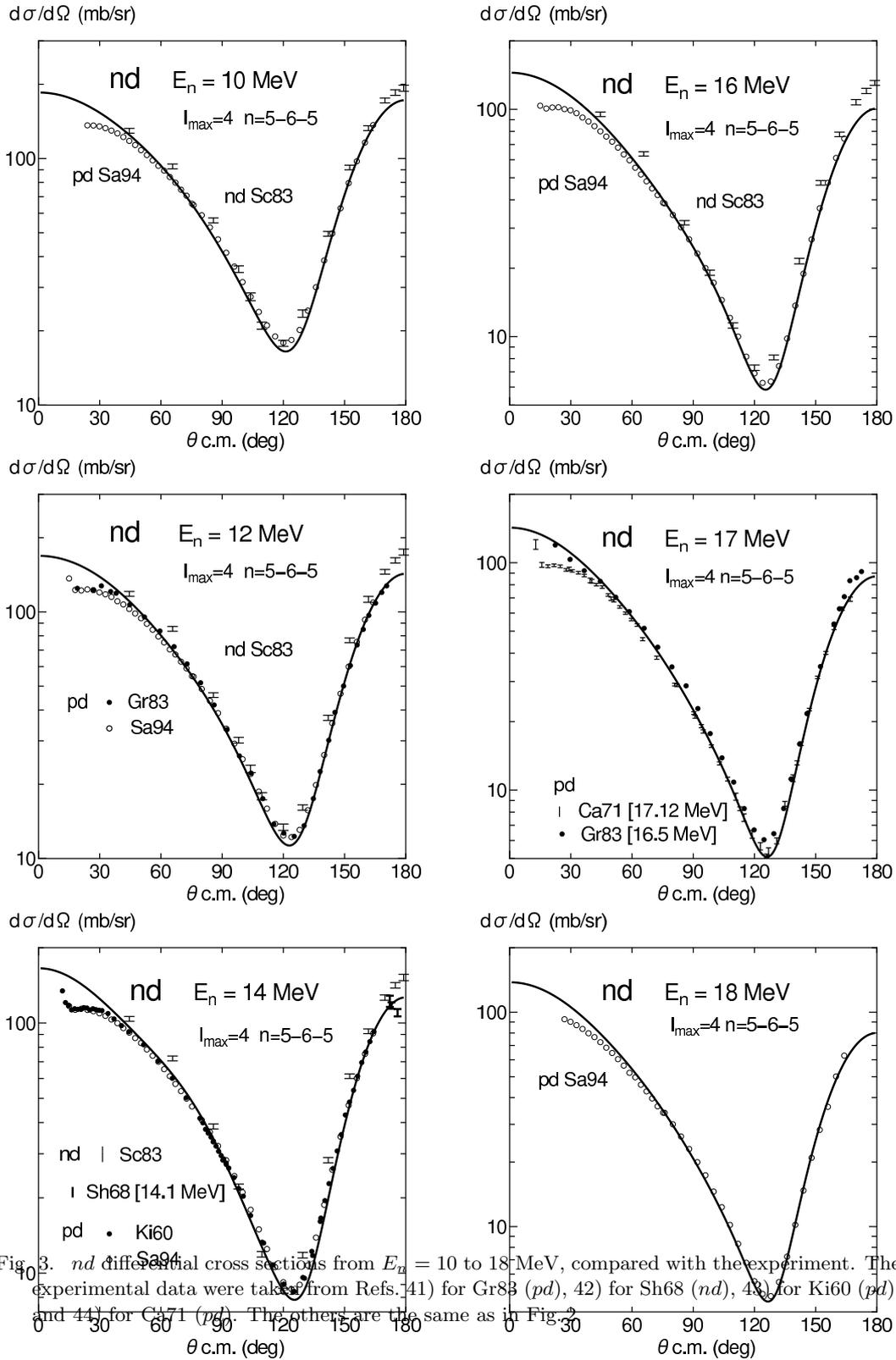}}
\vspace{-20mm}
\caption{$nd$ differential cross sections from $E_{n}=10$ to 18 MeV, compared with the experiment.
The experimental data were taken from Refs.~\citen{Gr83} for Gr83 ($pd$),
\citen{Sh68} for Sh68 ($nd$), \citen{Ki60} for Ki60 ($pd$), and \citen{Ca71} for Ca71 ($pd$).
The others are the same as in Fig.\,\ref{fig2}.}
\label{fig3}
\end{figure}

\begin{figure}[htb]
\centerline{\includegraphics[width=0.98\columnwidth]{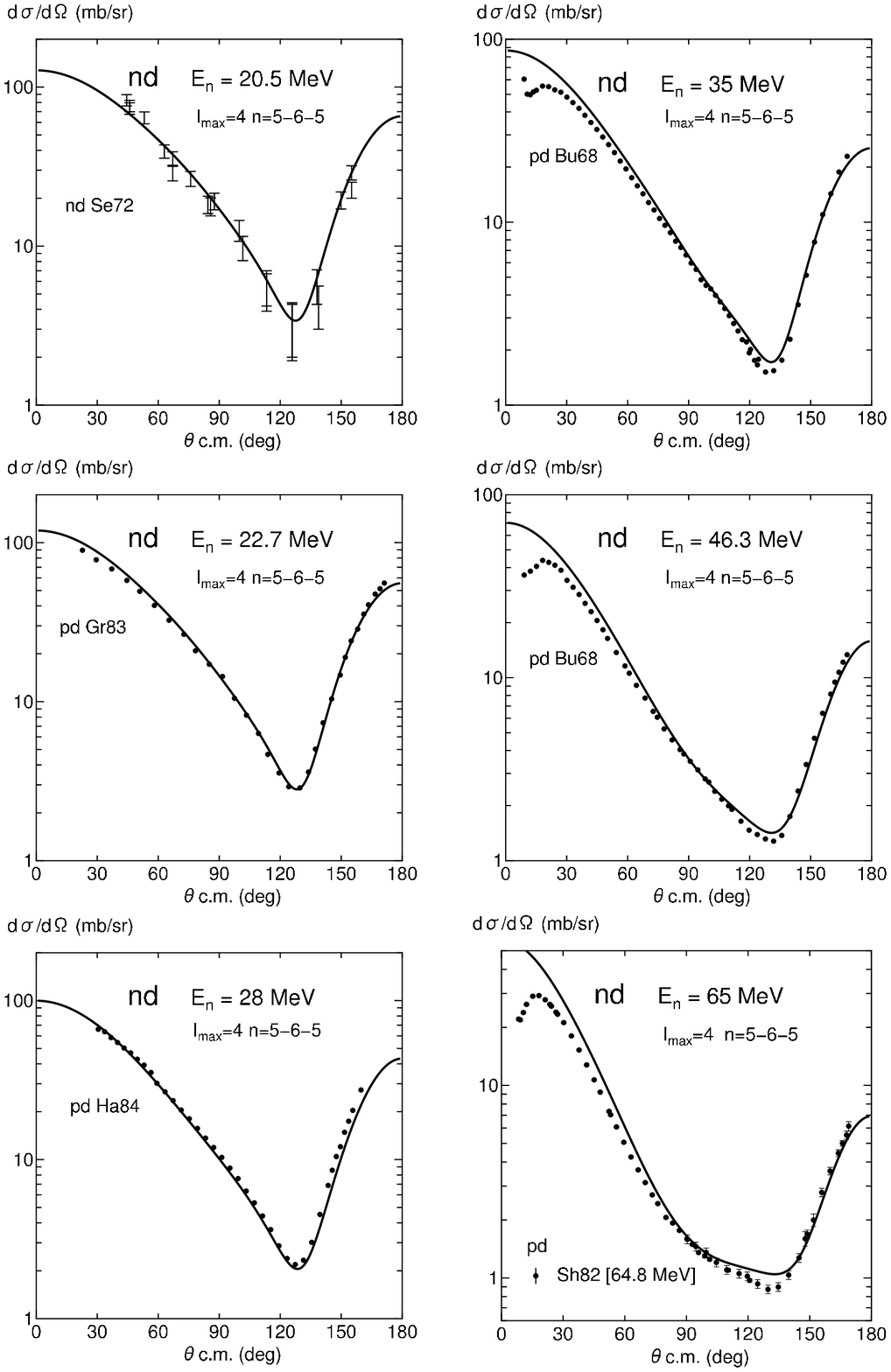}}
\vspace{-20mm}
\caption{$nd$ differential cross sections from $E_{n}=20.5$ to 65 MeV, compared with the experiment.
The experimental data were taken from Refs.~\citen{Ha84} for Ha84 ($pd$), \citen{Bu68} for Bu68 ($pd$),
and \citen{Sh82} for Sh82 ($pd$). The others are the same as in Figs.\,\ref{fig2} and \ref{fig3}.}
\label{fig4}
\end{figure}

We show in Figs.\,\ref{fig2} -- \ref{fig4} the $nd$ elastic differential cross sections predicted using fss2
for the neutron incident energies from $E_{n}=1$ to 65 MeV.
Here, we have used $I_{\rm max}=4$ and $n=5$-6-5 to obtain the well-converged values except for $E_{n} \leq 2$ MeV.
(The difference between $I_{\rm max}=2$ and $I_{\rm max}=4$ is shown in the panel of $E_{n}=3$ MeV as an example.)
These are compared with the $nd$ data plotted with bars. The experimental $pd$ differential cross sections
are also plotted with filled or open circles, unless otherwise specified.
For the energies $E_{n}=1$ and 2 MeV, we find that the $nd$ data and $pd$ data are fairly different from each other,
while this difference gradually diminishes for the energies $E_{n} \geq 3$ MeV
except for the forward angles $ \theta _{\rm cm}\leq 30^{\circ}$.
We can therefore compare our results with the $pd$ experimental data except for this angular region.
We find a satisfactory agreement with the experimental data.
In particular, the agreement around $E_{n}=17$ -- 22.7 MeV is excellent,
which is a common feature with the predictions using the meson-exchange potentials.\cite{PREP}
For higher energies $E_{n} \geq 35$ MeV, we find that the forward differential cross sections
are slightly overestimated in our model. It should be noted that in this energy region,
many partial waves contribute to \eq{dif1}, and yet the shape of the differential cross sections
is rather simple owing to the strong cancellation. The cross section minima (diffraction minima)
around $ \theta_{\rm cm}=120$ -- $130^{\circ}$ therefore afford a very crucial test of the two-nucleon interaction.
We find that the minimum values of the differential cross sections have an opposite energy dependence
to the one given by the standard meson-exchange potentials. 
This energy dependence is very important to discuss the effect of the three-nucleon force
for the meson-exchange potentials, which is generally known as the Sagara discrepancy \cite{Sa94} 
for the disagreement in the diffraction minima between experiment and theory.

\begin{table}[htb]
\caption{Comparison of the minimum values of the elastic differential cross sections
with the experimental $pd$ data after the Coulomb correction.
The minimum value $(d \sigma/d \Omega)_{\rm min}(nd)$ at the minimum point $ \theta _{\rm cm}$ is calculated 
for the $nd$ scattering at the neutron incident energy $E_{n}$.
The Coulomb correction, $[(pd)-(nd)]_{\rm AV18}$, for the difference between the $nd$ and $pd$ scattering is estimated
from the results in Table II of Ref.\,\citen{Ki01} for the AV18 potential,
and ``sum'' stands for our approximate prediction for the $pd$ scattering after the Coulomb correction.
The experimental data are taken from Refs.~\citen{Sa94} and \citen{Ki01}
for $E_{n} \leq 28$ MeV, Ref.~\citen{Bu68} for $E_{n}=35$ and 46.3 MeV,
and Ref.~\citen{Sh82} for $E_{n}=65$ MeV (exp: $E_{n}=64.8$ MeV.)}
\label{table1}
\renewcommand{\arraystretch}{1.2}
\setlength{\tabcolsep}{3mm}
\begin{center}
\begin{tabular}{cccccc}
\hline
$E_n$ & $\theta_{\rm cm}$ 
& $(d \sigma/d \Omega)_{\rm min}(nd)$
& $[(pd)-(nd)]_{\rm AV18}$ & sum
& $(d \sigma/d \Omega)^{\rm exp}_{\rm min}(pd)$ \\
(MeV) & (deg) & (mb) & (mb) & (mb) & (mb) \\ \hline
1  &  66 & 142.8 & 28.6 & 171.4 & $170.2 \pm 1.3$ \\
3  & 103 &  89.4 &  3.3 &  92.7 &  $91.1 \pm 0.7$ \\
5  & 112 &  50.3 &  3.1 &  53.4 &  $52.7 \pm 0.4$ \\
7  & 116 &  31.3 &  2.4 &  33.7 &  $32.9 \pm 0.2$ \\
9  & 120 &  20.3 &  1.8 &  22.1 &  $21.8 \pm 0.2$ \\
10 & 121 &  16.5 &  1.6 &  18.1 &  $18.0 \pm 0.2$ \\
12 & 123 &  11.3 &  1.0 &  12.3 &  $12.2 \pm 0.1$ \\
16 & 125 &  5.84 &  0.5 &   6.4 &   $6.2 \pm 0.1$ \\
18 & 127 &  4.47 &  0.4 &   4.9 &   $4.7 \pm 0.1$ \\
22.7 & 128 & 2.81 &  0.1 &  2.9 &   $2.89 \pm 0.03$ \\
28   & 129 & 2.06 &  0   &  2.1 &   $2.19 \pm 0.02$ \\
35   & 131 & 1.72 &      &      &   $1.52 \pm 0.04$ \\
46.3 & 132 & 1.43 &      &      &   $1.28 \pm 0.02$ \\
65   & 133 & 1.05 &      &      & $0.873 \pm 0.045$ \\ \hline
\end{tabular}
\end{center}
\end{table}

In order to investigate the energy dependence of the diffraction minima in more detail,
we have to incorporate the Coulomb force in our calculation since the precise data are only available for the $pd$ scattering.
Here, we estimate the Coulomb effect by using the published results
for the $nd$ and $pd$ cross sections in Ref.\,\citen{Ki01} for the AV18 potential.
Namely, we use the difference in the $nd$ and $pd$ cross section minima in Table II
and add it to our calculated results for the $nd$ scattering. The force dependence on the difference is
considered to be rather small, since the Coulomb force is a long-range force.
Table \ref{table1} shows such a comparison with the experimental $pd$ data.
We find that on the low-energy side with $E_{n} \leq 5$ MeV,
our estimated values reproduce the experimental data with an accuracy of less than 1 mb,
while on the high-energy side with $E_{n} \geq 35$ MeV, our results are slightly overestimated.
As for the low-energy side, we will show in a separate paper that the doublet scattering length $^{2}a$
of the low-energy $nd$ scattering is also consistently reproduced by fss2.\cite{KF10,SCL10}
These results are in accordance with the bound-state calculation of the triton,\cite{ren}
in which fss2 predicts a nearly correct binding energy close to the experiment without introducing the three-body force.
In the high-energy region with $E_{n} \geq 35$ MeV, it is reported that the Coulomb effect on the diffraction minima
is rather small and the inclusion of the $ \Delta $ isobar gradually becomes more important to increase them.\cite{De03,De05}
These observations imply a possibility that the rather large effect of the three-body force,
required for all the standard meson-exchange potentials, is related to the local form of the strong repulsive core,
introduced phenomenologically in the short-range region.
To confirm this, we need to investigate other $3N$ observables,
including the spin polarization and the deuteron breakup processes.
We have already obtained some good results, especially for the vector-analyzing power of the scattered
neutron,\cite{fb19,FK10} which we plan to report in a forthcoming paper.

\section{Summary}
We have applied our quark-model $NN$ interaction fss2 to the neutron-deuteron ($nd$) scattering
in the Faddeev formalism for systems of composite particles.
The energy dependence of the quark-model RGM kernel is eliminated by the standard off-shell transformation
utilizing the $1/\sqrt{N}$ factor, where $N$ is the normalization kernel for the two three-quark clusters.
This procedure yields an extra nonlocality, whose effect is very important to reproduce all the scattering observables
below $E_{n} \leq 65$ MeV. In this paper, we have developed our basic formulation
to solve the Alt-Grassberger-Sandhas (AGS) equations \cite{AGS} in the momentum representation,
using the off-shell RGM $t$-matrix generated from the energy-independent renormalized RGM kernel. 
The Gaussian nonlocal potential constructed from the fss2 is used in the isospin basis.\cite{apfb08}
The singularity of the $NN$ $t$-matrix from the deuteron pole is handled by the Noyes-Kowalski method.\cite{NO65,KO65}
Another notorious moving singularity of the free three-body Green function is treated by
the standard spline interpolation technique developed by the Bochum-Krakow group.\cite{spline82,Wi03,PREP,Liu05}
Together with the results in separate papers \cite{KF10,FK10,fb19,SCL10}
discussing the low-energy effective range parameters and the elastic scattering observables,
we have found many new features that seem to be related to the characteristic off-shell properties
possessed by the quark-model baryon-baryon interaction.
These include: 1) a large triton binding energy, 2) reproduction of the doublet scattering length $^{2}a$,
3) energy dependence of the diffraction minima of the differential cross sections, and
4) maximum height of the nucleon-analyzing power $A_{y}(\theta)$ in the low-energy region $E_{n} \leq 25$ MeV.  
Further investigations on the polarization observables and deuteron breakup processes will be
discussed in forthcoming papers. 

\section*{Acknowledgements}
The authors would like to thank Professor K. Miyagawa for giving them the main idea on
how to apply the spline interpolation method to the moving singularities.
They are indebted to Professors H. Witala, H. Kamada, and S. Ishikawa for many useful comments.
They also thank Professor K. Sagara for providing them with the $pd$ experimental data obtained by
the Kyushu university group. This work was supported by a Grant-in-Aid
for Scientific Research on Priority Areas (Grant No.~20028003) and
by a Grant-in-Aid for the Global COE Program
``The Next Generation of Physics, Spun from Universality and Emergence'' 
from the Ministry of Education, Culture, Sports, Science and Technology (MEXT) of Japan. 
It was also supported by the core-stage backup subsidies of Kyoto University.
The numerical calculations were carried out on Altix3700 BX2 at YITP in Kyoto University. 

\appendix
\section{Method to Calculate $\CK(\bp,\bp')$}
To calculate $W$ in \eq{qm4}, it is convenient to use the relationship
\begin{eqnarray}
W=\Lambda \left(\frac{1}{\sqrt{N}}h \frac{1}{\sqrt{N}}-h \right)\Lambda=\CK h+h\CK+\CK h \CK\ ,
\label{a1}
\end{eqnarray}
with
\begin{eqnarray}
\CK=\Lambda \left(\frac{1}{\sqrt{N}}-1 \right)\Lambda \ , 
\label{a2}
\end{eqnarray}
and calculate $\CK(\bp,\bp')$ in the momentum representation.
Since the Born kernel of $h=h_{0}+V_{\rm D}+G$ is already calculated,
we can easily obtain $W(\bp,\bp')$ by just a simple numerical integration.
In principle, the kernel $\CK$ is calculated from the power series expansion in $ \Lambda K \Lambda $ as
\begin{eqnarray}
\CK=\sum^{\infty}_{r=1}\frac{(2r-1)!!}{(2r)!!}\left(\Lambda K \Lambda \right)^{r}\ ,
\label{a3}
\end{eqnarray}
where the expansion formula
\begin{eqnarray}
\frac{1}{\sqrt{1-x}}-1=\sum^{\infty}_{r=1}\frac{(2r-1)!!}{(2r)!!}x^{r} \qquad \hbox{for} \qquad |x|<1\ ,
\label{a4}
\end{eqnarray}
is used. In the following, we will show that the infinite sum in \eq{a3} is taken analytically,
by using the power-series property for the eigenvalues of the exchange normalization kernel $K$.

We first consider, for simplicity, a single-channel system with only one quark (or nucleon) exchange
and at most one Pauli-forbidden state. The Pauli projection operator is $ \Lambda=1-|u_{00} \rangle
\langle u_{00}|$ with $u_{00}$ being the h.o. Pauli-forbidden state
defined in \eq{a12} below. The normalization kernel in the Bargmann space is expressed as
\begin{eqnarray}
N=e^{{\scriptsize \bz}^{*}\cdot{\scriptsize \bz'}}+X_{N}e^{\tau{\scriptsize \bz}^{*}\cdot{\scriptsize \bz'}}\ ,
\label{a5}
\end{eqnarray}
where $X_{N}$ is the spin-flavor (or spin-isospin) factor and $ \tau $ is given by $ \tau=1-1/\mu $ with
$ \mu $ being the reduced mass number. For the $n \alpha $ system, $X_{N}=-1$ and $ \tau=-1/4$.
For the $(3q)$--$(3q)$ system in the QM baryon-baryon interaction, $X_{N}$ is calculated numerically and $ \tau=1/3$.
We also have to consider the core exchange term in this case like in the $ \alpha \alpha $ system,
which will be discussed later. In the h.o.~basis, $K$ is expanded as
\begin{eqnarray}
K=-X_{N}e^{\tau{\scriptsize \bz}^{*}\cdot{\scriptsize \bz'}}=(-X_{N})\sum _{\scriptsize{\bN}}|\bN \rangle \tau^{N}\langle \bN|\ ,
\label{a6}
\end{eqnarray}
where $|\bN \rangle $ is the h.o.~states with the h.o.~quanta $\bN=(N_{x},N_{y},N_{z})$ and
$N=|\bN|=N_{x}+N_{y}+N_{z}=2n+\ell $ is the principal quantum number.
The eigenvalue of $K$ is given by $ \gamma _{N}=(-X_{N})\tau^{N}$($=(-1/4)^{N}$ for $n \alpha $).
The $r$th power of $K$ is easily calculated as
\begin{eqnarray}
K^{r} &=& (-X_{N})^{r}e^{\tau^{r}{\scriptsize \bz}^{*}\cdot{\scriptsize \bz'}}
=(-X_{N})^{r}\sum _{\scriptsize{\bN}}|\bN \rangle(\tau^{r})^{N}\langle \bN| \nonumber \\
&=& (-X_{N})^{r}\sum _{N \ell m}|N \ell m \rangle(\tau^{r})^{N}\langle N \ell m|\ .
\label{a7}
\end{eqnarray}
The kernel for $(\Lambda K \Lambda)^{r}$ can be obtained by restricting the sum in \eq{a7} over $N \geq 1$.

Suppose the spatial part of the GCM kernel for $K$ is $I_{N}(\bz;\bz')=e^{\tau{\scriptsize \bz}^{*}\cdot{\scriptsize \bz'}}$.
The corresponding Born kernel is given by
\begin{eqnarray}
M_{N}(\bq_{f},\bq_{i}) &=& \left(\frac{2 \pi}{\gamma}\frac{1}{1-\tau^{2}}\right)^{3/2}
\exp \left \{-\frac{1}{2 \gamma}\left(\frac{1-\tau}{1+\tau}\bq^{2}+\frac{1+\tau}{1-\tau}\frac{1}{4}\bk^{2}\right)\right \}
\nonumber \\
&=& \sum^{\infty}_{\ell=0}(2 \ell+1)M_{\ell}(q_{f},q_{i})P_{\ell}(\widehat{\bq}_{f}\cdot \widehat{\bq}_{i}) \nonumber \\
& = & 4 \pi \sum _{\ell m}M_{\ell}(q_{f},q_{i})Y_{\ell m}(\widehat{\bq}_{f})Y^{*}_{\ell m}(\widehat{\bq}_{i})\ ,
\label{a8}
\end{eqnarray}
where $\bk=\bq_{f}-\bq_{i}$, $\bq=(\bq_{f}+\bq_{i})/2$, and $ \gamma=\mu \nu $ with
$ \nu $ being the h.o.~width parameter of clusters. (See Appendix A of Ref.\,\citen{LSRGM}.)
Thus, for the partial wave component $K^{\rm space}_{\ell}=\sum^{\infty}_{n=0}|N \ell m \rangle
\tau^{N}\langle N \ell m|$ with $N=2n+\ell $, we find
\begin{eqnarray}
& & M_{\ell}(q,q';\tau)=\left(\frac{2 \pi}{\gamma}\frac{1}{1-\tau^{2}}\right)^{3/2}
\exp \left \{-\frac{1+\tau^{2}}{1-\tau^{2}}\frac{1}{4 \gamma}\left(q^{2}+q'^{2}\right)\right \}
~i_{\ell}\left(\frac{1}{\gamma}\frac{\tau}{1-\tau^{2}}qq'\right) \nonumber \\
& & =\left(\frac{2 \pi}{\gamma}\right)^{3/2}e^{-\frac{1}{4 \gamma}(q^{2}+q'^{2})}\left(\frac{1}{1-\tau^{2}}\right)^{3/2}
e^{-\frac{\tau^{2}}{1-\tau^{2}}\frac{1}{2 \gamma}(q^{2}+q'^{2})}i_{\ell}\left(\frac{1}{\gamma}\frac{\tau}{1-\tau^{2}}qq'\right)
\ , \label{a9}
\end{eqnarray}
where $i_{\ell}(x)$ is the imaginary spherical Bessel function. We use the notation
\begin{eqnarray}
i_{\ell}(x) &=& i^{\ell}j_{\ell}(-ix)=\frac{x^{\ell}}{(2 \ell+1)!!}F_{\ell}(x^{2})\ ,\nonumber \\
F_{\ell}(x) &=& \sum^{\infty}_{n=0}\frac{(2 \ell+1)!!}{(2n)!!(2n+2 \ell+1)!!}x^{n} \nonumber \\
&=& 1+\frac{1}{2 \cdot(2 \ell+3)}x+\frac{1}{2 \cdot 4 \cdot(2 \ell+3)(2 \ell+5)}x^{2}+\cdots \ .
\label{a10}
\end{eqnarray}
By using this notation, \eq{a9} can be expressed as
\begin{eqnarray}
M_{\ell}(q,q';\tau) &=& \tau^{\ell}~u_{0 \ell}(q)u_{0 \ell}(q')\left(\frac{1}{1-\tau^{2}}\right)^{\ell+3/2}
e^{-\frac{\tau^{2}}{1-\tau^{2}}\frac{1}{2 \gamma}(q^{2}+q'^{2})} \nonumber \\
& & \times F_{\ell}\left(\left(\frac{\tau}{1-\tau^{2}}\frac{qq'}{\gamma}\right)^{2}\right)\ ,
\label{a11}
\end{eqnarray}
where
\begin{eqnarray}
u_{0 \ell}(q)=u_{0 \ell}(q;\gamma)=\left(\frac{2 \pi}{\gamma}\right)^{3/4}
\frac{1}{\sqrt{(2 \ell+1)!!}}\left(\frac{q}{\sqrt{\gamma}}\right)^{\ell}e^{-\frac{q^{2}}{4 \gamma}}
\label{a12}
\end{eqnarray}
is the h.o.~wave function with $ \ell $ and the lowest h.o.~quanta $n=0$, normalized as
\begin{eqnarray}
\frac{4 \pi}{(2 \pi)^{3}}\int^{\infty}_{0}q^{2}dq~\left(u_{0 \ell}(q)\right)^{2}=1\ .
\label{a13}
\end{eqnarray}
Thus, we find
\begin{eqnarray}
K^{r}_{\ell}=(-X_{N})^{r}\sum^{\infty}_{n=0}\left(\tau^{r}\right)^{2n+\ell}u_{n \ell}(q)u_{n \ell}(q')
=(-X_{N})^{r}M_{\ell}(q,q';\tau^{r})\ .
\label{a14}
\end{eqnarray}
Here, we treat only the $n=0$ term separately. Namely, by defining a new function 
\begin{eqnarray}
\ \hspace{-10mm} \widetilde{M}_{\ell}(q,q';\tau)=\left(\frac{1}{1-\tau^{2}}\right)^{\ell+3/2}
e^{-\frac{\tau^{2}}{1-\tau^{2}}\frac{1}{2 \gamma}(q^{2}+q'^{2})}
F_{\ell}\left(\left(\frac{\tau}{1-\tau^{2}}\frac{qq'}{\gamma}\right)^{2}\right)-1\ ,\hfill
\label{a15}
\end{eqnarray}
we find
\begin{eqnarray}
K^{r}_{\ell} &=& (-X_{N}\tau^{\ell})^{r}u_{0 \ell}(q)u_{0 \ell}(q')\left[1+\widetilde{M}_{\ell}(q,q';\tau^{r})\right]\ .
\label{a16}
\end{eqnarray}
Thus, if there exists no Pauli-forbidden state, $ \Lambda=1$ and
\begin{eqnarray}
\CK_{\ell} &=& \frac{1}{\sqrt{N_{\ell}}}-1=\sum^{\infty}_{r=1}\frac{(2r-1)!!}{(2r)!!} K^{r}_{\ell} \nonumber \\
&=& \sum^{\infty}_{r=1}\frac{(2r-1)!!}{(2r)!!}(-X_{N}\tau^{\ell})^{r}u_{0 \ell}(q)u_{0 \ell}(q')
\left[1+\widetilde{M}_{\ell}(q,q';\tau^{r})\right]\ .
\label{a17}
\end{eqnarray}
For the first term, we take the $r$ sum with \eq{a4}. Then, we eventually obtain
\begin{eqnarray}
& & \CK_{\ell}(q,q')=u_{0 \ell}(q)u_{0 \ell}(q')\nonumber \\
& & \times \left[\frac{1}{\sqrt{1+X_{N}\tau^{\ell}}}-1+\sum^{\infty}_{r=1}\frac{(2r-1)!!}{(2r)!!}
(-X_{N}\tau^{\ell})^{r}\widetilde{M}_{\ell}(q,q';\tau^{r})\right].\hfill
\label{a18}
\end{eqnarray}
If a Pauli-forbidden state exists only for $(0s)$ with $ \ell=0$ (namely, $1+X_{N}=0$),
we have $X_{N}\tau^{\ell}=-1$ and the first term of \eq{a17} should be omitted. Namely,
\begin{eqnarray}
\CK_{0}(q,q') &=& \sum^{\infty}_{r=1}\frac{(2r-1)!!}{(2r)!!}\left(\Lambda K \Lambda \right)^{r}_{\ell=0} \nonumber \\
&=& u_{00}(q)u_{00}(q')\sum^{\infty}_{r=1}\frac{(2r-1)!!}{(2r)!!}\widetilde{M}_0(q,q';\tau^{r})\ .
\label{a19}
\end{eqnarray}

For $ \ell \neq 0$, the subtraction of one in $ \widetilde{M}_{\ell}(q,q';\tau)$ seems to be redundant
since the $(-1)$ term cancels with the first term in \eq{a16}. However, the convergence of
\begin{eqnarray}
\frac{1}{\sqrt{1+X_{N}\tau^{\ell}}}-1=\sum^{\infty}_{r=1}\frac{(2r-1)!!}{(2r)!!}\left(-X_{N}\tau^{\ell}\right)^{r}\ ,
\label{a20}
\end{eqnarray}
is very slow if $(-X_{N}\tau^{\ell})$ is close to 1. This happens in the $P$-wave case of the QM $NN$ interaction.
Namely, for the $NN$ interaction, we find (see Table II of Ref.\,\citen{Na95})
\begin{eqnarray}
\begin{array}{cccc}
\hbox{state} & X_{N} & \gamma _{N} & -X_{N}\tau^{\ell} \\ [2mm]
^{3}S_{1},\hbox{}^{1}S_{0} & \frac{1}{9} & -\frac{1}{9}\left(\frac{1}{3}\right)^{2n} & -\frac{1}{9} \\ [2mm]
^{1}P_{1} & -\frac{7}{3} & \frac{7}{3}\left(\frac{1}{3}\right)^{2n+1} & \frac{7}{9} \\ [2mm]
^{3}P_{1} & -\frac{31}{27} & \frac{31}{27}\left(\frac{1}{3}\right)^{2n+1} & \frac{31}{81} \\
\end{array}\ .
\label{a21}
\end{eqnarray}
On the other hand, $ \widetilde{M}_{\ell}(q,q';\tau)$ is a function of $ \tau^{2}$ with
$ \widetilde{M}_{\ell}(q,q';0)=0$. The leading term in the power series expansion of $ \tau^{2}$ is
\begin{eqnarray}
\widetilde{M}_{\ell}(q,q';\tau)=\tau^{2}\left[\ell+\frac{3}{2}-\frac{1}{2 \gamma}\left(q^{2}+q'^{2}\right)
+\frac{1}{2(2 \ell+3)}\left(\frac{qq'}{\gamma}\right)^{2}\right]+\cdots \ .
\label{a22}
\end{eqnarray}
Thus, we have $ \widetilde{M}_{\ell}(q,q';\tau)=O(\tau^{2})$ and $ \widetilde{M}_{\ell}(q,q';\tau^{r})=O(\tau^{2r})$.
Since $ \tau^{2}=1/9$ in the above example, increasing $r$ gives very fast convergence on the order of $1/10$ for $r=1$,
$1/100$ for $r=2$, $ \cdots $. We take the $r$ values up to $r_{\rm Max}=15$ in the actual calculation.

In the application to the $(3q)$--$(3q)$ RGM kernel, we need a proper treatment of the core exchange term
and the coupled-channel problem. In the operator formalism of the spin-flavor factors,
the basic $(3q)$--$(3q)$ GCM normalization kernel is expressed as
\begin{eqnarray}
I^{B}_{N}(\bz;\bz')=e^{{\bz}^{*}\cdot{\bz'}}
+X_{N}e^{\frac{1}{3}{\bz}^{*}\cdot{\bz'}}\ ,
\label{a23}
\end{eqnarray}
by which the full normalization kernel is given by
\begin{eqnarray}
N(\bz;\bz')=\frac{1}{2}\left[I^{B}_{N}(\bz;\bz')-P_{\sigma}P_{F}I^{B}_{N}(\bz;-\bz')\right]\ .
\label{a24}
\end{eqnarray}
Here, $P_{F}$ is the flavor exchange operator. Thus, the exchange normalization kernel $K$ is expressed as
\begin{eqnarray}
K(\bz;\bz')=-\frac{1}{2}\left(X_{N}~e^{\frac{1}{3}{\scriptsize \bz}^{*}\cdot{\scriptsize \bz'}}
-P_{\sigma}P_{F}~X_{N}~e^{-\frac{1}{3}{\bz}^{*}\cdot{\bz'}}\right)\ .
\label{a25}
\end{eqnarray}
The spin-flavor-color factor $X_{N}$ contains an explicit $P_{F}$ dependence, if the bra and ket sides
are composed of nonidentical baryons:
\begin{eqnarray}
X_{N}=X^{d}+X^{e}P_{\sigma}P_{F}\ .
\label{a26}
\end{eqnarray}
The exchange operator $P_{\sigma}P_{F}$ takes the value $-(-1)^{\ell}$,
where $(-1)^{\ell}$ is the parity of the two-baryon system. Thus, $X_{N}$ has an explicit parity dependence
\begin{eqnarray}
X^{\ell}_{N}=X^{d}-(-1)^{\ell}X^{e}\ ,
\label{a27}
\end{eqnarray}
which is important in actual calculations. The partial-wave component $K_{\ell}$ is then given by
\begin{eqnarray}
K_{\ell}=-X^{\ell}_{N}\sum _{N}|N \ell \rangle \left(\frac{1}{3}\right)^{N}\langle N \ell|\ .
\label{a28}
\end{eqnarray}
From now on, we omit the superscript $ \ell $ in $X^{\ell}_{N}$, by assuming a fixed $ \ell $.
The eigenvalue problem of the multichannel $K_{\ell}$ is reduced to the eigenvalue problem of
the matrix $(X_{N})_{\alpha \beta}$. We solve
\begin{eqnarray}
\sum _{\beta}\left(X_{N}\right)_{\alpha \beta}C^{\lambda}_{\beta}=\lambda C^{\lambda}_{\alpha}\ ,
\label{a29}
\end{eqnarray}
with 
\begin{eqnarray}
\sum _{\alpha}C^{\lambda}_{\alpha}C^{\lambda'}_{\alpha}=\delta _{\lambda,\lambda'}\ \ ,
\qquad \sum _{\lambda}C^{\lambda}_{\alpha}C^{\lambda}_{\beta}=\delta _{\alpha,\beta}\ .
\label{a30}
\end{eqnarray}
Then, the exchange norm kernel is given by
\begin{eqnarray}
(K_{\ell})_{\alpha \beta}=-\sum _{\lambda,N}\lambda C^{\lambda}_{\alpha}C^{\lambda}_{\beta}
\,|N \ell \rangle \left(\frac{1}{3}\right)^{N}\langle N \ell|\ .
\label{a31}
\end{eqnarray}
The full eigenvalue is $ \gamma _{\lambda N}=-\lambda(1/3)^{N}$. Only the $(0s)$ state is possible
for the Pauli-forbidden state in the isospin basis; i.e., the $SU_{3}$ $(11)_{s}$ state:
\begin{eqnarray}
\left(\Lambda K_{\ell}\Lambda \right)_{\alpha \beta}&=& \sum _{\lambda \left(\frac{1}{3}\right)^{N}\neq -1}
(-\lambda)\,C^{\lambda}_{\alpha}C^{\lambda}_{\beta}\,|N \ell \rangle \left(\frac{1}{3}\right)^{N} \langle N \ell|\ , \nonumber \\
\left(\left(\Lambda K_{\ell}\Lambda \right)^{r}\right)_{\alpha \beta} &=& \sum _{\lambda \left(\frac{1}{3}\right)^{N}\neq -1}
(-\lambda)^{r}\,C^{\lambda}_{\alpha}C^{\lambda}_{\beta}|N \ell \rangle \left(\frac{1}{3}\right)^{rN}\langle N \ell| \nonumber \\
&=& \left(\sum _{\lambda N}\left(-\lambda \right)^{r}\,C^{\lambda}_{\alpha}C^{\lambda}_{\beta}
\,|N \ell \rangle \left(\frac{1}{3}\right)^{rN} \langle N \ell|\right)
-\left(\delta _{\ell,0}\,C^{1}_{\alpha}C^{1}_{\beta}|00 \rangle \langle 00|\right) \nonumber \\
&=& \sum _{\lambda N}\left(K^{r}_{\ell}\right)_{\alpha \beta}-\delta _{\ell,0}C^{1}_{\alpha}C^{1}_{\beta}
|00 \rangle \langle 00|\ .
\label{a32}
\end{eqnarray}
The rest is almost the same as in the single-channel case. The final result is
\begin{eqnarray}
& & \left(\CK_{\ell}\right)_{\alpha \beta}=u_{0 \ell}(q)u_{0 \ell}(q')
\left[\sum _{\lambda \left(\frac{1}{3}\right)^{\ell}\neq -1}
\left(\frac{1}{\sqrt{1+\lambda \left(\frac{1}{3}\right)^{\ell}}}-1 \right)
C^{\lambda}_{\alpha}C^{\lambda}_{\beta} \right. \nonumber \\
& & \left. +\sum^{\infty}_{r=1}\frac{(2r-1)!!}{(2r)!!}\sum _{\lambda}\left(-\lambda \left(\frac{1}{3}\right)^{\ell}\right)^{r}
C^{\lambda}_{\alpha}C^{\lambda}_{\beta}~\widetilde{M}_{\ell}\left(q,q';\left(\frac{1}{3}\right)^{r}\right)
\right]\ .
\label{a33}
\end{eqnarray}

\section{Method to Calculate $Q_{k \mu \nu}$ in \eq{mv13}}
We first separate the integral region of \eq{mv13} as
\begin{eqnarray}
Q_{k \mu \nu}=(-2)\frac{1}{\omega _{\nu}}~\sum^{\kappa _{M}}_{\kappa=1}\int^{q_{\kappa}}_{q_{\kappa-1}}dq'
~Q_{k}(x_{0 \mu}+i0)S_{\nu}(q')\ ,
\label{b1}
\end{eqnarray}
with $q_{0}=0$ and $q_{\kappa _{M}}=q_{M}$, and apply the third-order spline function
\begin{eqnarray}
S^{(\kappa)}_{\nu}(q)=\sum^{3}_{m=0}\alpha^{\kappa(m)}_{\nu}~(q-q_{\kappa})^{m}
\qquad \hbox{for} \quad q \in[q_{\kappa-1},q_{\kappa}]\ .
\label{b2}
\end{eqnarray}
Then, we find
\begin{eqnarray}
Q_{k \mu \nu}=(-2)\frac{1}{\omega _{\nu}}~\sum^{\kappa _{M}}_{\kappa=1}\sum^{3}_{m=0}\alpha^{\kappa(m)}_{\nu}
~Q^{(k)}_{m \mu}(q_{\kappa-1},q_{\kappa})\ ,
\label{b3}
\end{eqnarray}
with
\begin{eqnarray}
Q^{(k)}_{m \mu}(q_{\kappa-1},q_{\kappa})=\int^{q_{\kappa}}_{q_{\kappa-1}}dq'
~Q_{k}(x_{0 \mu}+i0)(q'-q_{\kappa})^{m}\ .
\label{b4}
\end{eqnarray}
For $k=0$, a completely analytical calculation is possible by using the integral formula for logarithmic functions
\begin{eqnarray}
& & \int dx~{\rm log}(x+q)=(x+q)~[{\rm log}(x+q)-1]\ ,\nonumber \\
& & \int dx~x^{m}~{\rm log}x=\frac{1}{m+1}x^{m+1}\left[{\rm log}x-\frac{1}{m+1}\right]\ ,
\label{b5}
\end{eqnarray}
and
\begin{eqnarray}
& & I_{m}(a,b;q)\equiv \int^{b}_{a}dq'~{\rm log}\left|\frac{q'-q}{q'+q}\right|~(q'-b)^{m}
=(-1)^{m}\frac{1}{m+1}\left \{(b-q)^{m+1}~{\rm log}\left|\frac{b-q}{b+q}\right| \right. \nonumber \\
& & +\left[(b-a)^{m+1}-(b-q)^{m+1}\right]~{\rm log}\left|\frac{q-a}{q+a}\right|+\left[(b+q)^{m+1}-(b-q)^{m+1}\right]
~{\rm log}\left(\frac{q+a}{q+b}\right) \nonumber \\
& & \left. +\sum^{m-1}_{r=0}\frac{1}{r+1}\left[(b+q)^{m-r}-(b-q)^{m-r}\right]~(b-a)^{r+1}\right \}\ ,
\label{b6}
\end{eqnarray}
with $m=0,~1,~2,~\cdots $ and  $0 \leq a \leq b$. We can prove that $q \geq 0$ does not need to be
$0 \leq a \leq q \leq b$, but can also be at any position in the above form. The formula in \eq{b6}
is free from the logarithmic singularities since $ \lim _{\varepsilon \rightarrow 0}
~\varepsilon~{\rm log}\varepsilon=0$. We define $q_{\mu}$-dependent variables, $q_{1 \mu}$ and $q_{2 \mu}$,
for the crescent-shape region:
\begin{eqnarray}
\ \hspace{-10mm} & & \left \{\begin{array}{c}
q_{1 \mu}=\sqrt{\frac{3}{4}({q_{M}}^{2}-{q_{\mu}}^{2})}-\frac{1}{2}q_{\mu} \\
q_{2 \mu}=\sqrt{\frac{3}{4}({q_{M}}^{2}-{q_{\mu}}^{2})}+\frac{1}{2}q_{\mu} \\
\end{array} \right.\qquad \hbox{for} \quad q_{\mu}<\frac{\sqrt{3}}{2}q_{M}\ , \nonumber \\ [2mm]
& & \left \{\begin{array}{c}
q_{1 \mu}=\frac{1}{2}q_{\mu}-\sqrt{\frac{3}{4}({q_{M}}^{2}-{q_{\mu}}^{2})} \\
q_{2 \mu}=\frac{1}{2}q_{\mu}+\sqrt{\frac{3}{4}({q_{M}}^{2}-{q_{\mu}}^{2})} \\
\end{array} \right.\qquad \hbox{for} \quad \frac{\sqrt{3}}{2}q_{M}<q_{\mu}<q_{M}\ .
\label{b7}
\end{eqnarray}
Then, we obtain
\begin{eqnarray}
\ \hspace{-10mm}
& & Q^{(0)}_{m \mu}(q_{\kappa-1},q_{\kappa})=\left(-\frac{1}{2}\right)
\left[\pm I_{m}(q_{\kappa-1},q_{\kappa};q_{1 \mu})-I_{m}(q_{\kappa-1},q_{\kappa};q_{2 \mu}) \right. \nonumber \\
\ \hspace{-10mm} & & \qquad \left. +i \pi I^{\pi}_{m}(q_{\kappa-1},q_{\kappa};q_{1 \mu},q_{2 \mu})\right]
\qquad \hbox{for} \quad \left \{\begin{array}{c} q_{\mu}<\frac{\sqrt{3}}{2}q_{M} \\
\frac{\sqrt{3}}{2}q_{M}<q_{\mu}<q_{M} \\
\end{array} \right. \ ,
\label{b8}
\end{eqnarray}
where the theta function part is given by
\begin{eqnarray}
& & I^{\pi}_{m}(a,b;q_{1},q_{2})=\frac{1}{m+1}\left[\left({\rm Min}\{b,q_{2}\}-b \right)^{m+1}
-\left({\rm Max}\{a,q_{1}\}-b \right)^{m+1}\right] \nonumber \\ [2mm]
& & \hspace{20mm}\hbox{for} \quad {\rm Min}\,\{b,q_{2}\}>{\rm Max}\,\{a,q_{1}\}\ , \quad \hbox{otherwise} \quad 0\ .
\label{b9}
\end{eqnarray}

When $k \geq 1$, various methods are used to calculate $Q^{(k)}_{m \mu}(q_{\kappa-1},q_{\kappa})$ in \eq{b4} accurately.
First, the most accurate calculation outside the crescent area and its neighborhood is maybe the numerical integration
using the power series expansion of $Q_{k}(x)$ in $1/x^{2}$:
\begin{eqnarray}
Q_{k}(x)=\frac{2^{k}}{x^{k+1}}\sum^{\infty}_{n=0}\frac{(n+k)!(2n+k)!}{n!(2n+2k+1)!}\left(\frac{1}{x^{2}}\right)^{n}
\qquad \hbox{for} \quad x>1 \ .
\label{b10}
\end{eqnarray}
The convergence is so rapid that we can use \eq{b10} for the numerical integration
of $Q^{(k)}_{m \mu}(q_{\kappa-1},q_{\kappa})$ with $x_{0 \mu}(q'=q_{\kappa})>1.05$ or $x_{0 \mu}(q'=q_{\kappa-1})<-1.05$.
We use the 20-point Gauss-Legendre quadrature for the numerical integration.
(Note that there is no imaginary part appearing in this case.) In the crescent area and its neighborhood,
we will expand $P_{k}(x_{0 \mu})$ and $W_{k-1}(x_{0 \mu})$ in \eq{mv7} around $q'=q_{\kappa}$:
\begin{eqnarray}
& & P_{k}(x_{0 \mu})=\sum^{m_{M}}_{m'=0}\beta^{\kappa(m')}_{k \mu}~(q'-q_{\kappa})^{m'}\ , \nonumber \\
& & W_{k-1}(x_{0 \mu})=\sum^{m_{M}}_{m'=0}\beta^{W \kappa(m')}_{k \mu}~(q'-q_{\kappa})^{m'}\ ,
\label{b11}
\end{eqnarray}
for $q' \in [q_{\kappa-1},q_{\kappa}]$. Then, we can use the formulas in Eqs.\,(\ref{b6}) and (\ref{b9}) to obtain
\begin{eqnarray}
& & Q^{(k)}_{m \mu}(q_{\kappa-1},q_{\kappa})=\left(-\frac{1}{2}\right)\sum^{m_{M}}_{m'=0}
\left \{\beta^{\kappa(m')}_{k \mu}~\left[\pm I_{m+m'}(q_{\kappa-1},q_{\kappa};q_{1 \mu})
-I_{m+m'}(q_{\kappa-1},q_{\kappa};q_{2 \mu})\right. \right. \nonumber \\
& & \left. \left. +i \pi I^{\pi}_{m+m'}(q_{\kappa-1},q_{\kappa};q_{1 \mu},q_{2 \mu})\right]
+2 \beta^{W\,\kappa(m')}_{k \mu}~I^{\pi}_{m+m'}(q_{\kappa-1},q_{\kappa};q_{\kappa-1},q_{\kappa})\right \}\ ,
\label{b12}
\end{eqnarray}
for $q_{\mu}<(\sqrt{3}/2)q_{M}$ and $(\sqrt{3}/2)q_{M}<q_{\mu}<q_{M}$, respectively.
The expansion coefficients $ \beta^{\kappa(m')}_{k \mu}$ etc. are expressed using Bell's polynomials:\cite{Be34}
\begin{eqnarray}
Y_{0} &=& f_{0} \ ,\nonumber \\
Y_{1} &=& f_{1}~g_{1} \ ,\nonumber \\
Y_{2} &=& f_{1}~g_{2}+f_{2}~{g_{1}}^{2} \ ,\nonumber \\
Y_{3} &=& f_{1}~g_{3}+f_{2}~(3g_{2}~g_{1})+f_{3}~{g_{1}}^{3} \ ,\nonumber \\
Y_{4} &=& f_{1}~g_{4}+f_{2}~(4g_{3}~g_{1}+3{g_{2}}^2)+f_{3}~(6g_{2}~{g_{1}}^{2})+f_{4}~{g_{1}}^{4} \ ,~\cdots \ ,
\label{b13}
\end{eqnarray}
where the subscripts imply the higher derivatives. If we write \eq{b13} as $Y_{m}(f,g)$, 
$ \beta^{\kappa(m)}_{k \mu}$ etc. are expressed as
\begin{eqnarray}
& & \beta^{\kappa(m)}_{k \mu}=\frac{1}{m!}~Y_{m}\left(P_{k}(x_{0 \mu \kappa}),g_{\mu \kappa}\right)\ , \nonumber \\
& & \beta^{W\,\kappa(m)}_{k \mu}=\frac{1}{m!}~Y_{m}\left(W_{k-1}(x_{0 \mu \kappa}),g_{\mu \kappa}\right)\ .
\label{b14}
\end{eqnarray}
The higher derivatives of $g=g_{\mu \kappa}$ are given by
\begin{eqnarray}
g_{1}=-\left(\frac{x_{\mu}}{{q_{\kappa}}^{2}}+\frac{1}{q_{\mu}}\right)\ ,\quad
g_{2}=x_{\mu}\frac{2}{{q_{\kappa}}^{3}}\ ,\quad \cdots \ ,\quad
g_{r}=x_{\mu}(-1)^{r}\frac{r!}{{q_{\kappa}}^{r+1}}\ ,
\label{b15}
\end{eqnarray}
with $x_{\mu}=((3/4){q_{M}}^{2}-{q_{\mu}}^{2})/q_{\mu}$. For the practical calculation,
we first expand $P_{k}(x_{0 \mu})$ etc. around the middle point $(q_{\kappa-1}+q_{\kappa})/2$ and
then rearrange it to the form of \eq{b11}, by using $m_{M}=3$.
Actually, \eq{b12} cannot be used if $q_{\kappa}$ is small. This is because the higher derivative of $g$ in \eq{b15}
becomes very large for the small $q_{\kappa}$. This method is not valid either when $q_{\mu}$ is small
and $x_{0 \mu \kappa}$ rapidly changes from $-1$ to 1.
We therefore restrict the use of this method to the region with
$q_{\mu}>0.6~\hbox{fm}^{-1}$ and $q_{\kappa}>0.6~\hbox{fm}^{-1}$.

The third method to cover the above missing area is to use the separation of $Q_{k}(x+i0)$
to the singular and nonsingular parts:
\begin{eqnarray}
Q_{k}(x+i0)=\widetilde{Q}_{k}(x+i0)+P_{k}\left(\frac{x}{|x|}\right)~Q_{0}(x+i0)\ .
\label{b16}
\end{eqnarray}
Note that $ \widetilde{Q}_{0}(x+i0)=0$ and the nonsingular function $ \widetilde{Q}_{k}(x+i0)$
satisfies the same symmetry relation as $Q_{k}(x+i0)$, i.e.,
$ \widetilde{Q}_{k}(-x+i0)=(-1)^{k+1}{\widetilde{Q}_{k}(x+i0)}^{*}$ for real $x$. We calculate
\begin{eqnarray}
Q^{(k)}_{m \mu}(q_{\kappa-1},q_{\kappa})
&=& \int^{q_\kappa}_{q_{\kappa-1}}dq'~\widetilde{Q}_{k}(x_{0 \mu}+i0)(q'-q_{\kappa})^{m} \nonumber \\
& & +\int^{q_{\kappa}}_{q_{\kappa-1}}dq'~P_{k}\left(\frac{x_{0 \mu}}{|x_{0 \mu}|}\right)
~Q_{0}(x_{0 \mu}+i0)(q'-q_{\kappa})^{m}
\label{b17}
\end{eqnarray}
separately. The numerical integration is used for the first integral since the integrand is nonsingular.
However, if the interval $[q_{\kappa-1},q_{\kappa}]$ contains $q_{1 \mu}$ or $q_{2 \mu}$,
we separate the integral region into two parts, $[q_{\kappa-1},q_{1 \mu}]$ and $[q_{1 \mu},q_{\kappa}]$, etc.
The second integral in \eq{b17} is reduced to the previous formula, resulting in
\begin{eqnarray}
& & \int^{q_{\kappa}}_{q_{\kappa-1}}dq'~P_{k}\left(\frac{x_{0 \mu}}{|x_{0 \mu}|}\right)
~Q_{0}(x_{0 \mu}+i0)(q'-q_{\kappa})^{m} \nonumber \\
& & =\left \{\begin{array}{ll}
Q^{(0)}_{m \mu}(q_{\kappa-1},q_{\kappa}) & \hbox{for} \quad x_{0 \mu \kappa}>0 \\ [2mm]
(-1)^{k}~Q^{(0)}_{m \mu}(q_{\kappa-1},q_{\kappa}) & \hbox{for} \quad x_{0 \mu \kappa-1}<0 \\ [2mm]
Q^{(0)}_{m \mu}(q_{\kappa-1},q_{\kappa})+[(-1)^{k}-1]~Q^{(0)}_{m \mu}(q_{0 \mu},q_{\kappa}) &
\hbox{for} \quad x_{0 \mu \kappa}<0<x_{0 \mu \kappa-1} \\ \end{array} \right. \ ,\nonumber \\
\label{b18}
\end{eqnarray}
for $q_{\mu}<(\sqrt{3}/2)q_{M}$, where $Q^{(0)}_{m \mu}(a,b)$ is obtained from \eq{b8}
by replacing $q_{\kappa-1}$ with $a$ and $q_{\kappa}$ with $b$.
In the last term in \eq{b18}, we have defined $q_{0 \mu}=\sqrt{(3/4){q_{M}}^{2}-{q_{\mu}}^{2}}$.
In the case of $(\sqrt{3}/2)q_{M}<q_{\mu}<q_{M}$, only the second case of \eq{b18} is realized.

\end{document}